\documentclass{emulateapj}
\usepackage{apjfonts}
\usepackage{graphicx}
\usepackage{amsmath}
     \usepackage{dpfloat}
     \usepackage{multirow}

\usepackage{longtable}

\usepackage[toc,page]{appendix}

\usepackage{color}

\def\kms{\ifmmode{\rm km\thinspace s^{-1}}\else km\thinspace s$^{-1}$\fi}

\shortauthors{Rappaport et al.~2016}
\shorttitle{Interacting Quadruple Star System}

\begin{document}

%
\def\ltsima{$\; \buildrel < \over \sim \;$}
\def\lsim{\lower.5ex\hbox{\ltsima}}
\def\gtsima{$\; \buildrel > \over \sim \;$}
\def\gsim{\lower.5ex\hbox{\gtsima}}
\def\teff{$T_\mathrm{eff}$}                 
 \def\vsini{\hbox{$v$\,sin\,$i_{\star}$}}    
  \def\cms2{\hbox{\,cm\,s$^{-2}$}}

%

\bibliographystyle{apj}

\title{EPIC 220204960: A Quadruple Star System Containing Two Strongly Interacting Eclipsing Binaries}

\author{
S.~Rappaport\altaffilmark{1}, 
A.~Vanderburg\altaffilmark{2},
T.~Borkovits\altaffilmark{3},
B.~Kalomeni\altaffilmark{1,4},
J.P.~Halpern\altaffilmark{5},
H.~Ngo\altaffilmark{6},
G.~N. Mace\altaffilmark{7},
B.J.~Fulton\altaffilmark{8},
A.W.~Howard\altaffilmark{9},
H.~Isaacson\altaffilmark{10},
E.A.~Petigura\altaffilmark{11},
D.~Mawet\altaffilmark{9},
M.H.~Kristiansen\altaffilmark{12,13},
T.L.~Jacobs\altaffilmark{14},
D.~LaCourse\altaffilmark{15},
A.~Bieryla\altaffilmark{2},
E.~Forg\'acs-Dajka\altaffilmark{16},
L.~Nelson\altaffilmark{17}
}

\altaffiltext{1}{Department of Physics, and Kavli Institute for Astrophysics and Space Research, Massachusetts Institute of Technology, Cambridge, MA 02139, USA, sar@mit.edu}
  
\altaffiltext{2}{Harvard-Smithsonian Center for Astrophysics, 60 Garden Street, Cambridge, MA 02138 USA; avanderburg@cfa.harvard.edu}

\altaffiltext{3}{Baja Astronomical Observatory of Szeged University, H-6500 Baja, Szegedi \'{u}t, Kt. 766, Hungary; borko@electra.bajaobs.hu}

\altaffiltext{4}{Department of Astronomy and Space Sciences, Ege University, 35100, \.Izmir, Turkey; belinda.kalomeni@ege.edu.tr}

\altaffiltext{5}{Department of Astronomy, Columbia University, New York, NY; jules@astro.columbia.edu}

\altaffiltext{6}{California Institute of Technology, Division of Geological and Planetary Sciences, 1200 E California Blvd MC 150-21, Pasadena,
CA 91125, USA; hngo@caltech.edu}

\altaffiltext{7}{McDonald Observatory and the Department of Astronomy, The University of Texas at Austin, Austin, TX 78712, USA; gmace@astro.as.utexas.edu}

\altaffiltext{8}{Institute for Astronomy, University of Hawai'i, 2680 Woodlawn Drive, Honolulu, HI 96822, USA; bfulton@hawaii.edu}  

\altaffiltext{9}{Astronomy Department, California Institute of Technology, MC 249-17, 1200 E. California Blvd., Pasadena, CA 91125, USA} 

\altaffiltext{10}{Department of Astronomy, University of California at Berkeley, Berkeley, CA, 94720-3411, USA} 

\altaffiltext{11}{Hubble Fellow, Astronomy Department, California Institute of Technology, Pasadena, California, USA} 

\altaffiltext{12}{DTU Space, National Space Institute, Technical University of Denmark, Elektrovej 327, DK-2800 Lyngby, Denmark}

\altaffiltext{13}{Brorfelde Observatory, Observator Gyldenkernes Vej 7, DK-4340 T\o ll\o se, Denmark}

\altaffiltext{14}{12812 SE 69th Place Bellevue, WA 98006}  

\altaffiltext{15}{7507 52nd Place NE Marysville, WA 98270} 

\altaffiltext{16}{Astronomical Department, E\"otv\"os University, H-1118 Budapest, P\'azm\'any P\'eter stny. 1/A, Hungary} 

\altaffiltext{17}{Department of Physics and Astronomy, Bishop's University, 2600 College St., Sherbrooke, QC J1M 1Z7}

\slugcomment{{\it Monthly Notices of the Royal Astronomical Society}, Accepted 2016 January 17}

\begin{abstract}

We present a strongly interacting quadruple system associated with the K2 target EPIC 220204960.  The K2 target itself is a $K_p = 12.7$ magnitude star at $T_{\rm eff} \simeq 6100$ K which we designate as ``B-N'' (blue northerly image).  The host of the quadruple system, however, is a $K_p \simeq 17$ magnitude star with a composite M-star spectrum, which we designate as ``R-S'' (red southerly image).  With a 3.2$''$ separation and similar radial velocities and photometric distances, `B-N' is likely physically associated with `R-S', making this a quintuple system, but that is incidental to our main claim of a strongly interacting quadruple system in `R-S'.  The two binaries in `R-S' have orbital periods of 13.27 d and 14.41 d, respectively, and each has an inclination angle of $\gtrsim 89^\circ$.  From our analysis of radial velocity measurements, and of the photometric lightcurve, we conclude that all four stars are very similar with masses close to $0.4 \, M_\odot$.  Both of the binaries exhibit significant ETVs where those of the primary and secondary eclipses `diverge' by 0.05 days over the course of the 80-day observations.  Via a systematic set of numerical simulations of quadruple systems consisting of two interacting binaries, we conclude that the outer orbital period is very likely to be between 300 and 500 days.  If sufficient time is devoted to RV studies of this faint target, the outer orbit should be measurable within a year.  

\end{abstract}


\section{Introduction}

Higher-order multiple star systems are interesting to study for several reasons.  Such systems (i) provide insights into star-formation processes; (ii) allow for a study of short-term (i.e., $\lesssim$ few years) perturbative dynamical interactions among the constituent stars; and (iii) enable us to learn more about longer-term dynamical interactions that can actually alter the configuration of the system (e.g., via Kozai-Lidov cycles; Kozai 1962; Lidov 1962).  These multi-component stellar systems can be discovered, studied, and tracked via a wide variety of techniques including historical photographic plates (e.g., Frieboes-Conde \& Herczeg 1973; Borkovits \& Heged\"us 1996), searches for common proper motion stellar systems (e.g., Raghavan et al.~2012); ground-based photometric monitoring programs searching for gravitational microlensing events (MACHO; e.g., Alcock et al.~2000; OGLE; e.g.,  Pietrukowicz et al.~2013) or planet transits (e.g., SuperWASP, Lohr et al.2015a; HATNet, Bakos et al.~2002; KELT, Pepper et al.~2007), high-resolution imaging or interferometric studies (e.g., Tokovinin 2014a, 2014b), and spectroscopy aimed at measuring radial velocities (Tokovinin 2014a). 

Perhaps the quickest pathway to discovering close multiple interacting star systems is via the study of eclipsing binaries whose eclipse timing variations (`ETVs') indicate the presence of a relatively nearby third body or perhaps even another binary.  In a series of papers based on precision {\em Kepler} photometry (see, e.g., Borucki et al.~2010; Batalha et al.~2011), some 220 triple-star candidates were found via their ETVs (Rappaport et al.~2013; Conroy et al.~2014; Borkovits et al.~2015; Borkovits et al.~2016).  Several of the {\em Kepler} binary systems turned out to be members of quadruple systems consisting of two gravitationally bound binaries (KIC 4247791: Lehmann et al.~2012; KIC 7177553: Lehmann et al.~2016: and quintuple EPIC 212651213: Rappaport et al.~2016).  One of the {\em Kepler} systems, KIC 4150611/HD 181469, is arranged as a triple system bound to two other binaries (Shibahashi \& Kurtz 2012, and references therein; Prsa et al.~2016).  

Other interesting quadruple star systems include: 1SWASP J093010.78+533859.5 (Lohr et al.~2015b); the young B-star quintuple HD 27638 (Torres 2006); HD 155448 (Sch\"utz et al.~2011); 14 Aurigae (Barstow et al.~2001); $\sigma^2$ Coronae Borealis (Raghavan et al.~2009); GG Tau (Di Folco et al.~2014); and HIP 28790/28764 and HIP 64478 (Tokovinin 2016). 

Perhaps the two quadruples in a binary-binary configuration (i.e., `2+2') with the shortest known outer periods are V994 Her (1062 days; Zasche \& Uhla\v{r} 2016) and VW LMi (355 days; Pribulla et al.~2008). $\xi$-Tau (145 days; Nemravova et al.~2016) is a quadruple in a `2+1+1' configuration which puts it in a somewhat different category.  The scale of dynamical perturbations of one binary by the other can be characterized by the parameter: $P_{\rm bin}^2/P_{\rm out}$, where $P_{\rm bin}$ and $P_{\rm out}$ are the binary and outer period, respectively.  The values of this quantity are 0.004 d and 0.18 d for V994 Her and VW LMi, respectively.  The value of this parameter for $\xi$-Tau, where the binary is largely perturbed by a single star, is 0.35 d.  

In this work we report the discovery with K2 of a strongly interacting quadruple system consisting of two eclipsing binaries, with orbital periods of 13.27 d and 14.41 d and all four M stars having very similar properties. Both binaries exhibit strong ETVs from which we infer an outer period of $\sim$ a year that, in turn, implies $P_{\rm bin}^2/P_{\rm out} \approx 0.54 \,(P_{\rm out}/{\rm yr})^{-1} $ d.  Such a substantial value of this parameter could turn out to be the largest among the known sample of quadruples.

This work is organized as follows.  In Sect.~\ref{sec:K2} we describe the 80-day K2 observation of EPIC 220204960 with its two physically associated eclipsing binaries.  Our ground-based observations of the two stellar images associated with this target are presented in Sect.~\ref{sec:ground}.  These include classification spectra and Keck AO imaging.  In Sect.~\ref{sec:RVs} we discuss the six radial-velocity spectra that we were able to obtain, and the resultant binary orbital solutions. The discovery of significant and substantial ETVs in the eclipses of both binaries are presented in Sect.~\ref{sec:ETVs}.  We use a physically-based model to evaluate the eclipsing binary lightcurves in Sect.~\ref{sec:MCMC}, and thereby determine many of the system parameters of the binaries not available from the radial velocities, as well as independent mass determinations.  In Sect.~\ref{sec:lcfactory} we re-introduce a method for simultaneously modeling the two eclipsing binary lightcurves, and the results are compared with those derived in Sect.~\ref{sec:MCMC}.  In Sect.~\ref{sec:numerical} we simulate via numerical integrations the dynamical interactions of the four stars in the quadruple system, and set substantial constraints on the outer period of the two binaries orbiting each other.  The dynamical perturbations of each binary on the other are assessed analytically in Sect.~\ref{sec:analytic}.  We summarize our results and draw some final conclusions in Sect.~\ref{sec:concl}.

\vspace{0.6cm}

\section{K2 Observations}
\label{sec:K2}

As part of our ongoing search for eclipsing binaries, we downloaded all available K2 Extracted Lightcurves common to Campaign 8 from the MAST\footnote{\url{http://archive.stsci.edu/k2/data\_search/search.php}}. We utilized both the Ames pipelined data set and that of Vanderburg \& Johnson (2014).  The flux data from all 24,000 targets were searched for periodicities via Fourier transforms and the BLS algorithm (Kov\'acs et al.~2002). The folded lightcurves of targets with significant peaks in their FFTs or BLS transforms were then examined by eye to look for unusual objects among those with periodic features.  In addition, some of us (MHK, DL, and TLJ) visually inspected all the K2 light curves for unusual stellar or planetary systems.  

Within a few days after the release of the Field 8 data set, EPIC 220204960 was identified as a potential quadruple star system by both visual inspection and via the BLS algorithmic search.  After identifying four sets of eclipses in the K2 light curve, we re-processed the light curve by simultaneously fitting for long-term variability,  K2 roll-dependent systematics, and the four eclipse shapes in the light curves using the method described in Vanderburg et al. (2016). For the rest of the analysis, we use this re-processed light curve and divide away the best-fit long-term variability, since it was dominated by an instrumental trend. 

\begin{figure}[t]
\begin{center}
\includegraphics[width=0.48 \textwidth]{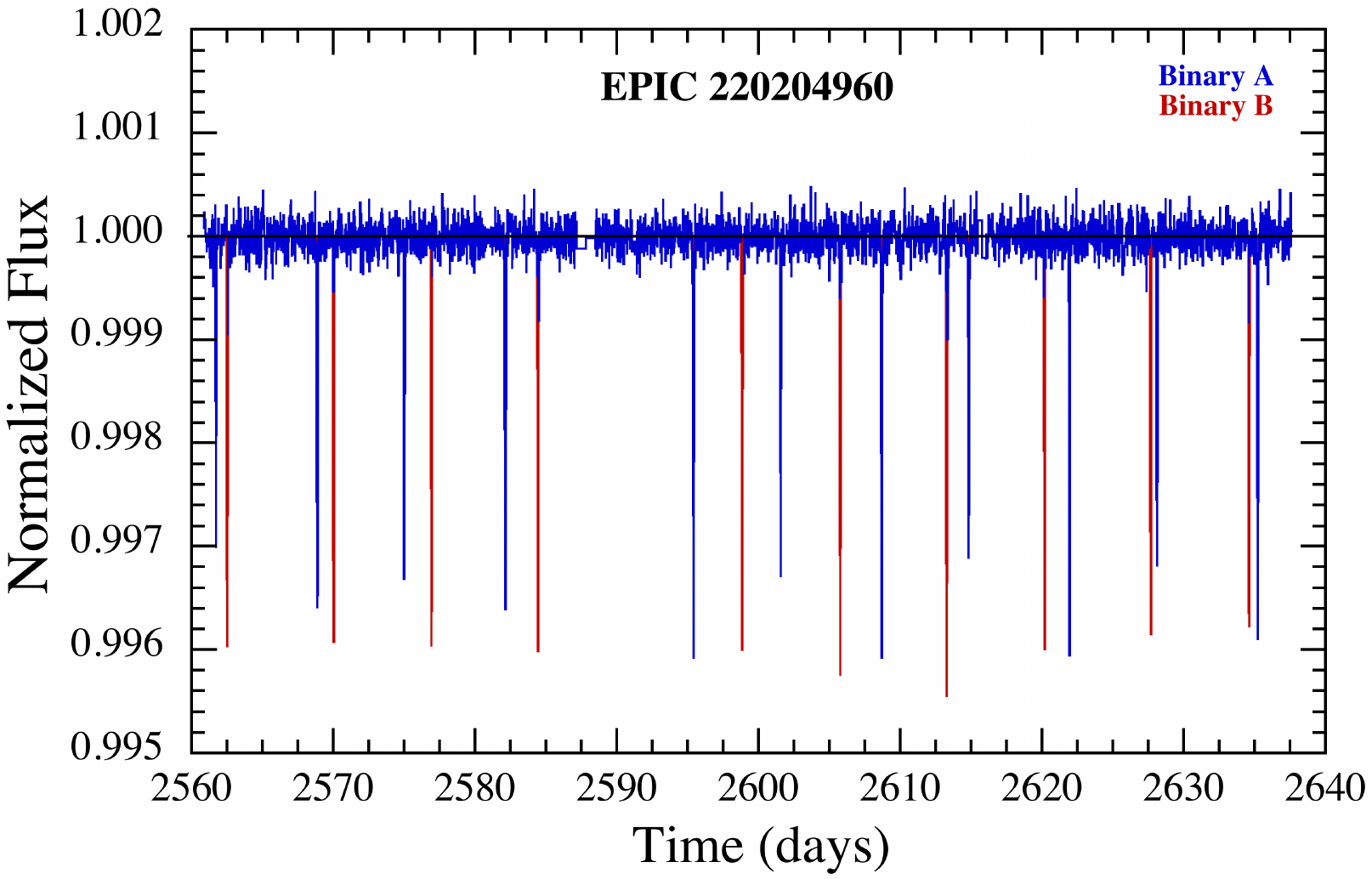}
\caption{K2 flux data for EPIC 220204960.  The eclipses of the 13.27-day `A' binary are colored in blue, while those of the 14.41-day `B' binary are in red.  All four eclipses are of comparably shallow depth.  We note that this lightcurve contains the light of the bright northerly blue stellar image designated `B-N' (see Fig.~\ref{fig:SDSS}).  At $\sim$3$''$ separation from the `A' and `B' binaries, the fluxes are not separable with K2.}
\label{fig:rawLC}
\end{center}
\end{figure}

The basic lightcurve is shown in Fig.~\ref{fig:rawLC}, where three features are obvious by inspection.  (1) All four eclipses of the two binaries have very similar depths, though the secondary eclipse in the A binary has about 3/4 the depth of the primary.   (2) The periods of the two binaries are quite comparable with $P_A = 13.27$ d and $P_B = 14.41$ d.  (3) The eclipse depths are remarkably shallow at $\sim$0.4\%.  We rather quickly inferred that the coincidence of the similar sets of extraordinarily shallow eclipses indicates a dilution effect from a neighboring star, rather than two precisely inclined orbits that happen to produce such tiny eclipse depths.  Quantitatively, we note that for eclipsing binaries with two similar stars the a priori probability of an undiluted eclipse of 0.4\% is only $\sim$0.02.  The probability of this occurring by chance in two related binaries is only $5 \times 10^{-4}$.  

The primary and secondary eclipses in both binaries are close to being equally spaced, but are measurably different from being equal.  We define the fractional separations between eclipses as, $\Delta t_{s,p}/P_{\rm orb} =(t_{\rm sec}-t_{\rm pri})/P_{\rm orb}$, where $t_{\rm sec}$ and $t_{\rm pri}$ are times of sequential secondary and primary eclipses, and $t_{\rm sec} > t_{\rm pri}$.  The fitted fractional separations between the two eclipses are: $0.4633 \pm 0.0001$ and $0.4797 \pm 0.0001$, for the A and B binaries, respectively. We can then utilize the approximate expression (good to 2nd order in eccentricity $e$):
\begin{equation}
e\,\cos \omega \simeq \frac{\pi}{2} \left[\frac{\Delta t_{s,p}}{P_{\rm orb}} - \frac{1}{2} \right]
\label{eqn:ecom}
\end{equation}
where $\omega$ is the argument of periastron of the primary component (derived from a Taylor series expansion of Eqn.~\ref{eqn:Ddef}; from Sterne 1939), to say that $e \cos \omega_A \simeq -0.0577$ and $e \cos \omega_B \simeq -0.0319$, for the A and B binaries, respectively.  

We can also utilize information from the relative widths of the two eclipses, $w_1$ and $w_2$, to find a measure of $e\,\sin \omega$.  For small $e$ and arbitrary $\omega$:
\begin{equation}
e\,\sin \omega \simeq \frac{(1-w_{\rm pri}/w_{\rm sec})}{(1+w_{\rm pri}/w_{\rm sec})}
\label{eqn:esom}
\end{equation}
(see, e.g., Kopal 1959, Chapt.\,VI).  From the K2 photometry, we determine that $w_{\rm A,pri}/w_{\rm A,sec} = 1.13 \pm 0.05$, and $w_{\rm B,pri}/w_{\rm B,sec} = 1.09 \pm 0.04$.  Therefore, $e_A \, \sin \omega_A = -0.061 \pm 0.023$ and  $e_B \, \sin \omega_B = -0.042 \pm 0.020$.  Thus, based on the limits obtained from Eqns.~(\ref{eqn:ecom}) and (\ref{eqn:esom}) we can constrain the orbital eccentricities and arguments of periastron of the A and B binaries to be
$$0.058 \lesssim e_A \lesssim 0.10 ~~ {\rm and} ~~ 0.032 \lesssim e_B \lesssim 0.07 $$
$$ \omega_A \simeq 230^{+10}_{-30} ~{\rm deg}~~ {\rm and} ~~\omega_B \simeq 240^{+10}_{-40} ~ {\rm deg}$$
Thus, not only are the binaries very similar in other respects, they both have small, but distinctly non-zero eccentricities. 

We return to a more detailed quantitative analysis of the lightcurves of the two binaries in Sections \ref{sec:ETVs}, \ref{sec:MCMC}, and \ref{sec:lcfactory}.

\begin{table}
\centering
\caption{Properties of the EPIC 220204960 System}
\begin{tabular}{lcc}
\hline
\hline
Parameter &
220204960 `B-N' &
220204960 `R-S' \\
\hline
RA (J2000) & 00:48:32.65  & 00:48:32.67 \\  
Dec (J2000) &  00:10:18.59 & 00:10:15.20 \\  
$K_p$ & 12.66 & ... \\
$u^a$ &  15.08  & 24.64 \\
$B^b$ & 13.31 & ...\\
$g^a$ & 13.02 & 18.01 \\
$G^b$ & 12.58 & 16.82 \\
$V^b$ & 12.76 & ... \\
$R^b$ & 12.63 & ... \\
$r^a$ & 12.71 & 16.44 \\  
$z^a$ & 13.37 & 15.51 \\
$i^b$ & 12.54 & ... \\
J$^c$ & 11.75 & 14.2 \\
H$^c$ & 11.54 & ... \\
K$^c$ & 11.44 & 13.4 \\
W1$^d$ & 11.28 & ... \\
W2$^d$ & 11.30 & ... \\
W3$^d$ & 11.40 & ... \\
W4$^d$ & ... & ... \\
Distance (pc)$^e$ & $560 \pm 150$ &  $600 \pm 150$ \\   
$\mu_\alpha$ (mas ~${\rm yr}^{-1}$)$^f$ & $-0.1 \pm 1.3$ & ... \\ 
$\mu_\delta$ (mas ~${\rm yr}^{-1}$)$^f$ &  $-8.5 \pm 1.4$ & ... \\ 
\hline
\label{tbl:mags}
\end{tabular}

{\bf Notes.} (a) Taken from the SDSS image (Ahn et al.~2012). (b) From VizieR \url{http://vizier.u-strasbg.fr/}; UCAC4 (Zacharias et al.~2013). (c) 2MASS catalog (Skrutskie et al.~2006).  (d) WISE point source catalog (Cutri et al.~2013). (e) Based on photometric parallax only.  This utilized adapted V magnitudes of 12.76 and 17.1 for the two stellar images, the bolometric luminosities for the four M stars given in Table \ref{tbl:MCMC}, the bolometric magnitude of the `B-N' image inferred from Table \ref{tbl:BN}, and appropriate bolometric corrections for the M stars in question. (f) From UCAC4 (Zacharias et al.~2013); Smart \& Nicastro (2014); Huber et al.~(2015).
\end{table}

\vspace{0.5cm}

\section{Ground Based Observations}
\label{sec:ground}

\subsection{SDSS Image}
\label{sec:SDSS}
  
\begin{figure}[h]
\begin{center}
\includegraphics[width=0.48 \textwidth]{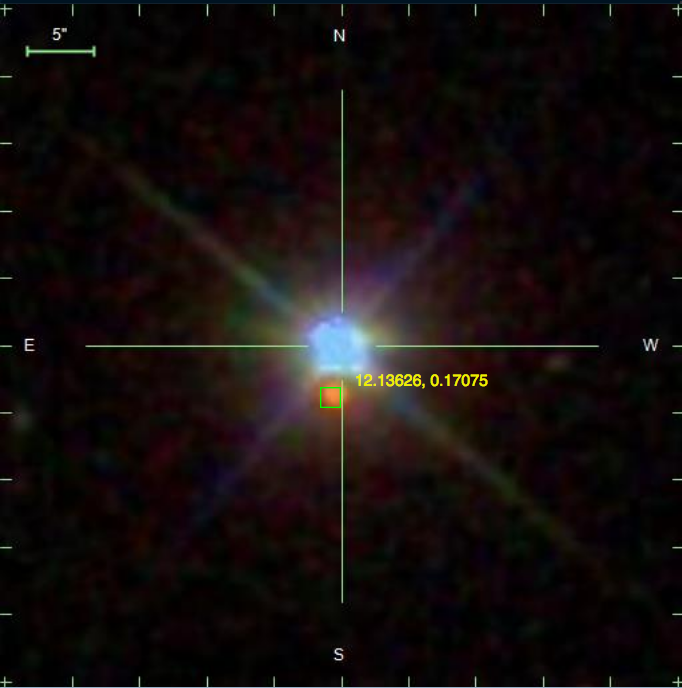}
\caption{SDSS image showing the region near EPIC 220204960.  We have designated the brighter bluish colored image to the north as `B-N' while the fainter reddish image some 3$''$ to the south is designated as `R-S'.  The `R-S' image hosts both binaries in a bound quadruple system.}
\label{fig:SDSS}
\end{center}
\end{figure} 

The SDSS image of EPIC 220204960 is shown in Fig.~\ref{fig:SDSS}.  The brighter bluish image to the north (hereafter `B-N') dominates the light, but note the fainter reddish image some 3$''$ to the south (hereafter `R-S').  We summarize the available properties of these two stars in Table \ref{tbl:mags}.  

Through the {\em Kepler} bandpass, the `R-S' image ranges from between 2.8 and 5 magnitudes fainter than the `B-N' image.  When we carefully integrate these magnitudes, as well as our detailed spectra (see Sect. \ref{sec:MDM}), more quantitatively over the {\em Kepler} bandpass,  we find a flux ratio of $45 \pm 10$ (90\% confidence) between the `B-N' and `R-S' images.  As we will show, this difference is sufficient to explain the extreme dilution of the eclipses provided that both binaries are hosted within the `R-S' image.

\subsection{MDM Spectra}
\label{sec:MDM}
  
On 2016 August 31 UT, two 1500-s spectra of EPIC 220204960 were obtained with the
Ohio State Multi-Object Spectrograph (OSMOS) on the 2.4\,m Hiltner telescope of the
MDM Observatory on Kitt Peak, Arizona.  In long-slit mode, a $1\farcs2$ slit was
aligned with the two stellar images for the first exposure.  The second exposure
had the slit oriented east-west through image `R-S'.  A volume phase holographic
grism provided a dispersion of 0.72~\AA\ pixel$^{-1}$ and a resolution of 2.9 \AA\
on a Silicon Technology Associates STA-0500 CCD with $4064\times4064$ $15\,\mu$ pixels.
The wavelength coverage is 3967--6876~\AA. The dispersion solution was derived
from 28 comparison lines of Hg and Ne, yielding rms residuals of 0.02~\AA, although a
systematic error of up to 0.4~\AA\ could be present due to instrument flexure.

\begin{figure}[h]
\begin{center}
\includegraphics[width=0.49 \textwidth]{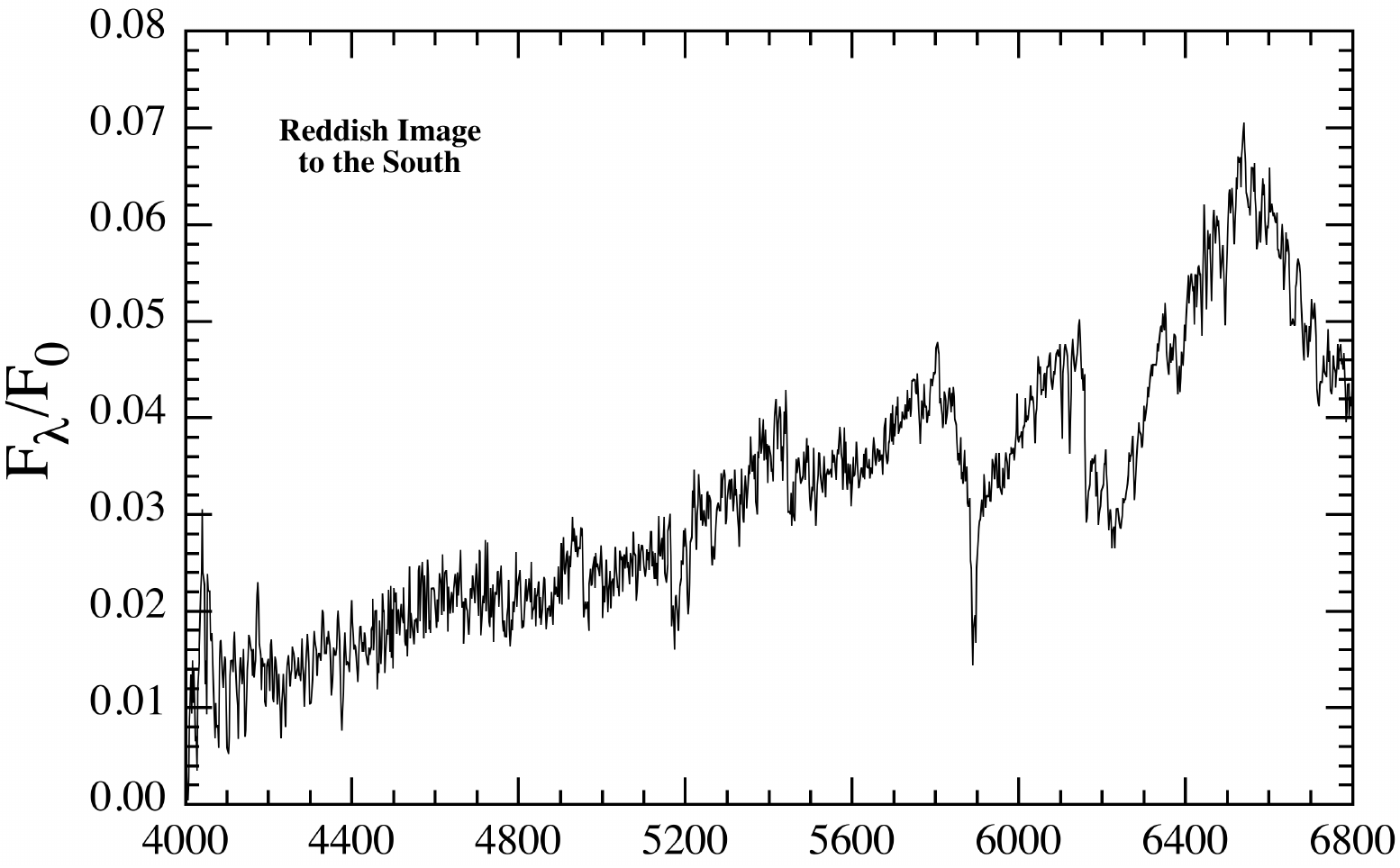} 
\includegraphics[width=0.49 \textwidth]{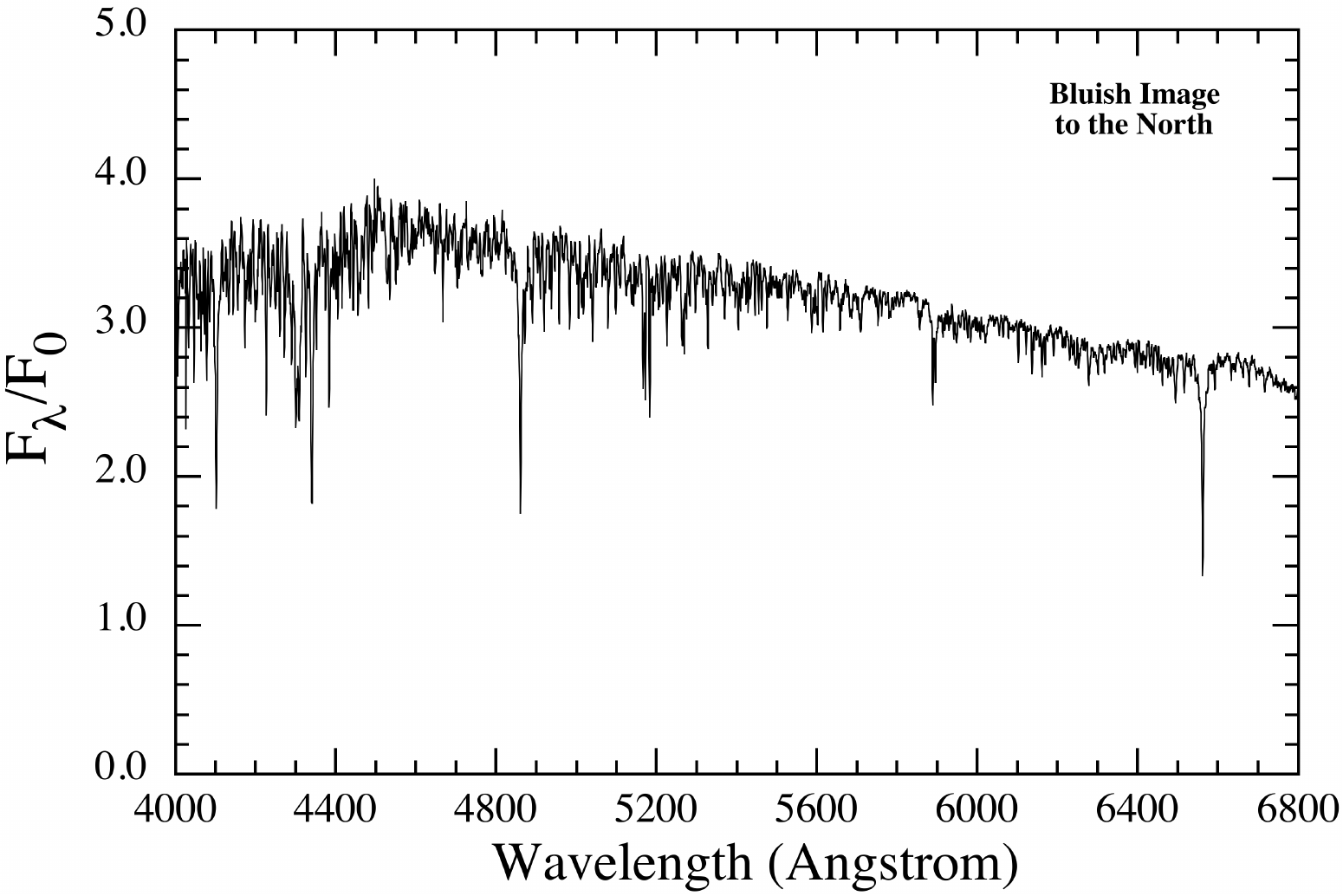} 
\caption{MDM 2.4-m spectra of the `R-S' image (top panel) and `B-N' star (bottom panel).  The spectra have been corrected for the throughput efficiency as a function of wavelength.  The reference flux density, $F_0$, is $10^{-14}$ ergs cm$^{-2}$ s$^{-1}$ \AA \,$^{-1}$. The ratio of detected flux in the two spectra is $\sim$100.  This implies a ratio of $\sim$60 in the {\em Kepler} bandpass after correcting for the red flux between 6800 \AA \, and 8500 \AA \, that is not included in the spectrum.}
\label{fig:MDM}
\end{center}
\end{figure}

The spectra for both the `B-N' image and `R-S' image are shown in Fig.~\ref{fig:MDM}.
The east-west slit was used here to extract the spectrum of `R-S,' as it had less
contamination from `B-N'.  There is no detectable leakage of the spectrum of
`B-N' into `R-S', as the prominent Balmer absorption lines in `B-N' are absent in `R-S.'  
Although the narrow slit and sky conditions were not conducive to absolute
spectrophotometry, the standard star HD 19445 was used for flux calibration.
The equivalent slit magnitude of `B-N' is $V\approx12.6$, in reasonable agreement
with the value in Table~\ref{tbl:mags} ($V=12.76$).

It is clear that the spectrum of `R-S' is that of an early M star. Examining the Pickles (1998) atlas
of stellar spectra, we find a best match with an M2.5V type.  Although,
it is worth noting that this is actually a composite spectrum of four,
very likely similar, stars.  By contrast, the `B-N' image is that of a G2V star.

\subsection{Spectral Classification of the `B-N' Image from TRES Spectrum}
\label{sec:TRES}

\begin{figure*} 

\begin{center}
\includegraphics[width=0.80 \textwidth]{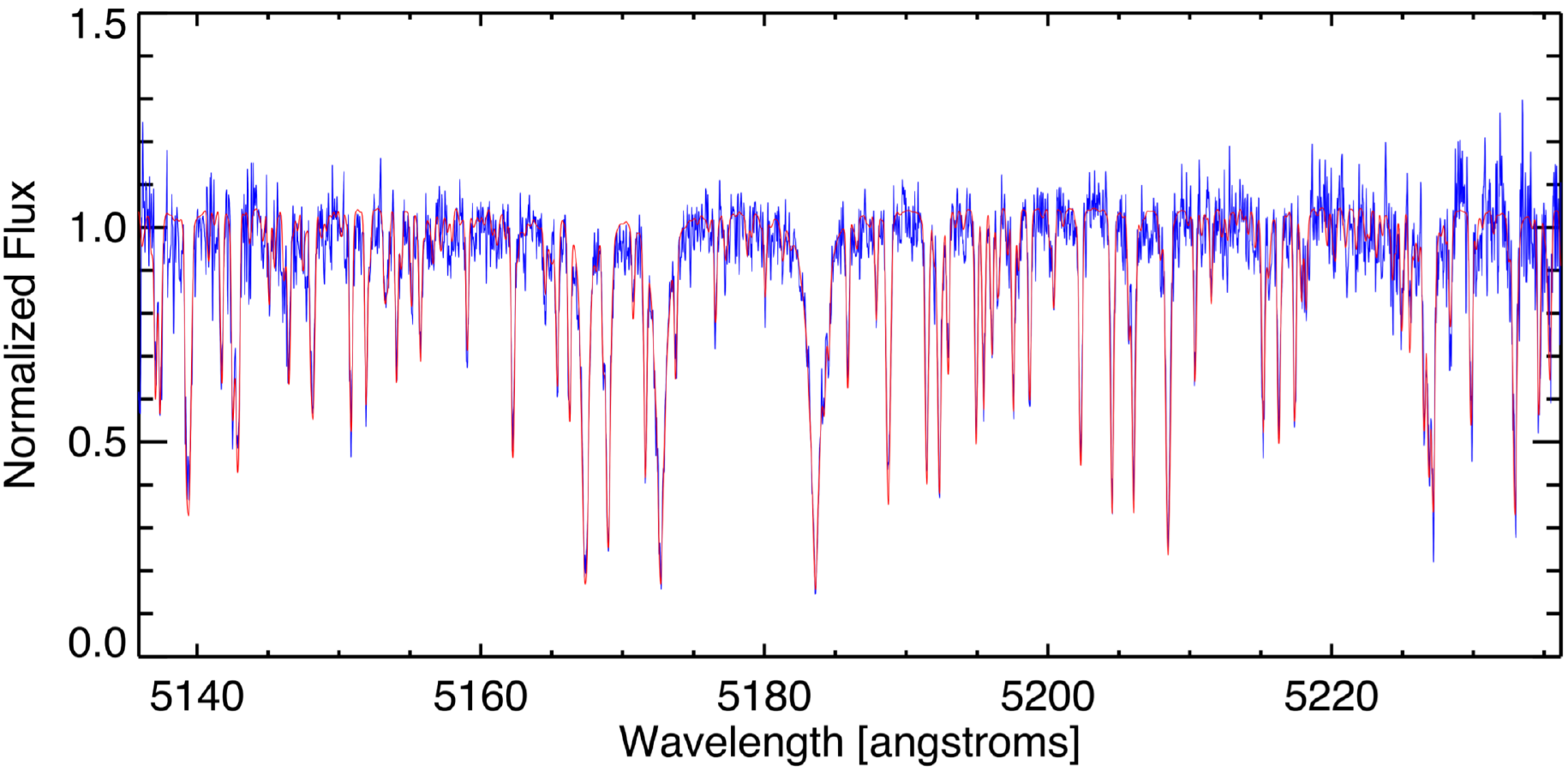} 
\caption{200 \AA \, segment of the overall TRES spectrum of EPIC 220204960 used to characterize the `B-N' image.  Data are plotted in blue while the fitted model curve is shown in red.  The results of the model fit are summarized in Table \ref{tbl:BN}.}
\label{fig:TRES}
\end{center}
\end{figure*}

We observed the blue northern component of EPIC 220204960 with the Tillinghast Reflector Echelle Spectrograph (TRES) on the 1.5 meter telescope on Mt.~Hopkins, AZ. 1500-s and 2000-s exposures were taken on 2016 July 13 UT and 2016 Oct.~24 UT, respectively.  These yielded spectra with signal-to-noise ratio of $\sim$30 per resolution element at 520 nm, and a spectral resolving power of R = 44,000.  We reduced the spectra following Buchhave et al.~(2010). A portion of one spectrum is shown in Fig.~\ref{fig:TRES}.  We measured an absolute radial velocity for the `B-N' image of EPIC 220204960 by cross-correlating the observed TRES spectrum against a suite of synthetic model spectra based on Kurucz (1992) model atmospheres.  The velocities for the two measurements were $-4.505$ and $-4.516$ km s$^{-1}$, consistent with no change at $11 \pm 50$ m s$^{-1}$.  These have been corrected for the gravitational blueshift to the barycenter.  They also have a residual, systematic, error (in common) of 100 m s$^{-1}$.


\begin{figure}
\begin{center}
\includegraphics[width=0.40 \textwidth]{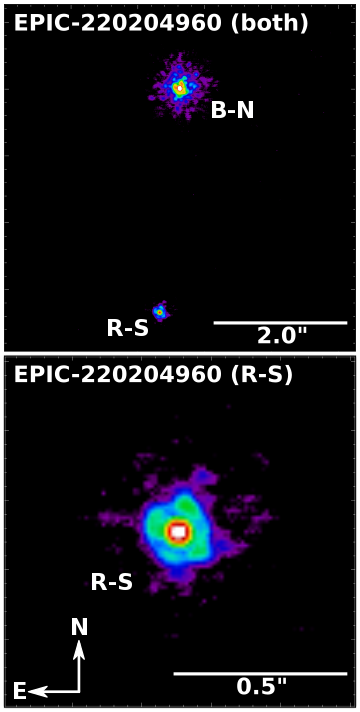}
\caption{Top panel: Keck-AO image in $K_s$-band of EPIC 220204960, including the brighter blue image to the north, `B-N', and the fainter red image $3\farcs4$ to the south, `R-S'.  A zoom-in on the `R-S' image which hosts the quadruple system, is shown in the bottom panel.  If the two binaries were separated by $0\farcs1$ or more, the core of the image would be cleanly split into two objects. A separation of even $0\farcs05$ would produce a noticeably elongated central core, which is not seen. }
\label{fig:AO}
\end{center}
\end{figure}  

We measured the stellar parameters of the `B-N' image using the Stellar Parameter Classification code (SPC, Buchhave et al.~2010, 2012). SPC cross correlates an observed spectrum against a grid of synthetic spectra based on Kurucz atmospheric models (Kurucz 1992). The analysis yielded $T_{\rm eff}=6085 \pm 72$ K, $\log \, g = 4.23 \pm 0.02$, [m/H] = $0.16 \pm 0.13$, and $v\, \sin\,i=7.6 \pm 0.2$ km s$^{-1}$ (see Table \ref{tbl:BN}).   

\subsection{Adaptive Optics Imaging}
\label{sec:Keck_AO}

We obtained natural guide star observations of both the `B-N' and `R-S' components of EPIC 220204960 on 2016 July 19 UT to better characterize this quadruple system. We used the narrow camera setting (10 milliarcseconds, mas, per pixel) of the NIRC2 camera (PI: Keith Matthews) on Keck II. We used dome flat fields and dark frames to calibrate the images and remove artifacts. 

We acquired 12 frames of EPIC 220204960 in each of the $J$ and $K_s$ bands (central wavelengths of 1.250 $\mu$m and 2.145 $\mu$m, respectively) for a total on-sky integration time of 240 seconds in each band. Figure \ref{fig:AO} shows a stacked $K_s$ band image of both components of this target (cf.~the SDSS image in Fig.~\ref{fig:SDSS}).  The northerly `B-N' image is separated by $3\farcs359\pm0\farcs002$ from the southerly red image `R-S' at a position angle of $174.60 \pm 0.03$ degrees east of north.  Photometry and $K_s$ band astrometry were computed via PSF fitting using a combined Moffat and Gaussian PSF model following the techniques described in Ngo et al.~(2015) and the NIRC2 distortion solution presented in Service et al.~(2016).  The `B-N' component is $2.43\pm 0.03$ magnitudes brighter than `R-S' in the $K_s$ band ($2.50 \pm 0.01$ magnitudes in $J$).  The fact that the `B-N'/`R-S' flux ratio is only $\sim$10 in the NIR, compared to $\sim$45 in the {\em Kepler} band, is an indication of how red the `R-S' image is.

The evidence presented in the next section shows that both binaries are actually hosted by the `R-S' image. In the bottom panel of Fig.~\ref{fig:AO}, we show a zoomed-in image of the `R-S' component.  This blown-up image looks distinctly single, and shows no sign of the core even being elongated.  We have carried out simulations of close pairs of comparably bright images, at a range of spacings, and we conclude from this that separations between the two binaries of $\gtrsim 0.05''$ can be conservatively ruled out.  At a source distance of some 600 pc, this sets an upper limit on the projected physical separation of $\sim$30 AU.

\begin{table}
\centering
\caption{Properties of Stellar Image `B-N'}
\begin{tabular}{lc}
\hline
\hline
Parameter & Value \\
\hline
$T_{\rm eff}$ [K]$^a$ \ &  $6085 \pm 72$  \\ 
$\log ~g$  [cgs]$^a$ &  $4.23 \pm 0.02$ \\
$M$ [$M_\odot$]$^b$ &  $1.20 \pm 0.07$  \\ 
$R$ [$R_\odot$]$^b$ & $1.35 \pm 0.18$  \\
$L$ [$L_\odot$]$^b$ & $2.3 \pm 0.7$ \\
$\gamma$ [km/s]$^a$ & $-4.510 \pm 0.062$  \\
$v \, \sin \,i$ [km/s]$^a$ & $7.6 \pm 0.2$ \\
$[m/H]$$^a$  & $0.16 \pm 0.13$ \\
$\mathcal{F}_{BN}/\mathcal{F}_{RS}$$^c$ & $45 \pm 10$ \\
\hline
\end{tabular}
\label{tbl:BN}

{\bf Notes.} (a) Taken from the analysis of two TRES spectra acquired on 2016 July 13 and 2016 Oct.~24 (see Sect.~\ref{sec:TRES}).  (b) Derived from $T_{\rm eff}$ and $\log~g$ using the Yonsei--Yale tracks (Yi et al.~2001) for an assumed solar composition.  (c) Based on the MDM spectra (see Sect.~\ref{sec:MDM}), and the magnitudes given in Table \ref{tbl:mags}.
\end{table}

\section{A Few Radial Velocity Measurements}
\label{sec:RVs}

Because the `R-S' image, which hosts all four M stars, is relatively faint, we have been able to obtain only six spectra at five independent epochs of the quality required for radial velocity measurements.  Two were taken with the IRGINS spectrograph mounted on the Discovery Channel Telescope, while four others were acquired with the HIRES spectrometer on Keck.  By coincidence, the second of the two IGRINS spectra was taken within three hours of the first of the HIRES spectra, and therefore these nearly simultaneous spectra serve as a consistency check between the two sets of data.

\subsection{IGRINS Spectra}
\label{sec:IGRINS}

The Immersion Grating Infrared Spectrometer (IGRINS) employs a silicon immersion grating for broad spectral coverage at high-resolution in the near-infrared. The design provides high throughput and an unprecedented R $\approx$ 45,000 spectrum of both the H and K bands (1.45-2.5 $\mu$m). IGRINS was initially commissioned on the 2.7\,m Harlan J. Smith Telescope at McDonald Observatory (Park et al.~2014; Mace et al.~2016) before being deployed to the Discovery Channel Telescope (DCT) in September 2016. The `R-S' image was observed once during IGRINS commissioning at the DCT on UT 2016 Sept.~19 and again during regular science operation on UT 2016 Oct.~10. These observations were taken in ABBA nod sequences with 900\,s and 1200\,s exposure times. The spectra were optimally extracted using the IGRINS Pipeline Package (Lee \& Gullikson 2016). Dome-flats were taken at the start of the night and wavelengths were determined using sky lines. Telluric correction by A0V stars at similar air masses to EPIC 220204960 provide a flattened spectrum with a signal-to-noise of 30-40 per resolution element. The longer exposure times required for this fainter target resulted in higher OH residuals in the spectrum from 2016 Oct.~10.

\subsection{HIRES Spectrum}
\label{sec:HIRES}

We observed the red southern component of EPIC 220204960 with the High Resolution Echelle Spectrometer (HIRES, Vogt 1994) on the Keck I telescope on Mauna Kea. We used the standard California Planet Search observing setup with the red cross disperser and the C2 0\farcs86 decker (Howard et al.~2010). We obtained 20-minute exposures on 2016 Oct.~10, Nov.~21, and Nov.~26 and a 15-minute exposure on Nov.~5, yielding signal-to-noise ratios that were typically between 5 and 20 per pixel between 500 and 800 nm. 

\begin{figure}
\begin{center}
\includegraphics[width=0.48 \textwidth]{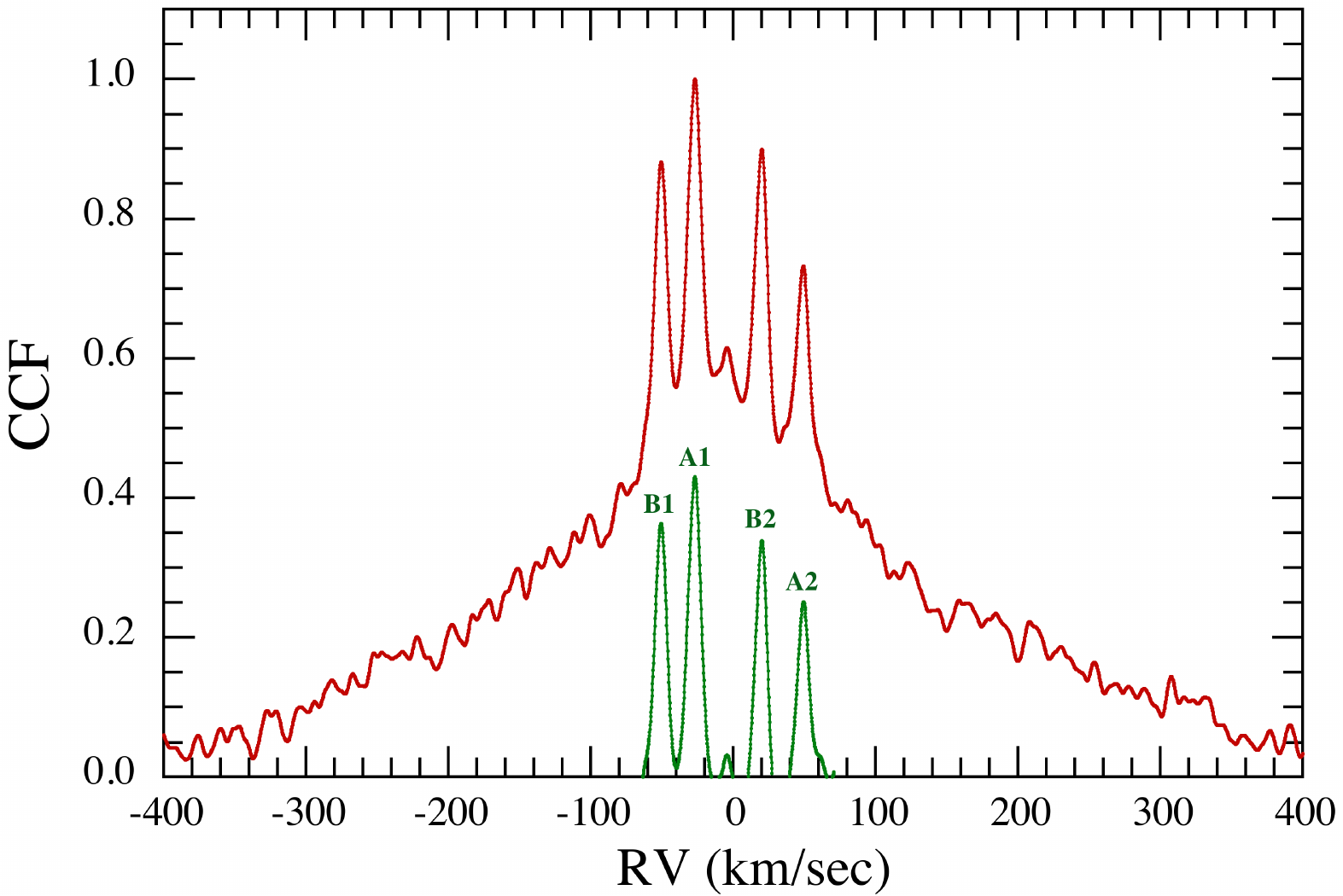}
\caption{Cross correlation (red curve) between the HIRES spectrum and the template from a reference M star.  After subtracting off a pedestal of broader features, the green curve shows the four peaks more clearly that correspond to the four M stars in the quadruple star system. The inferred radial velocities, which range from $-50$ km s$^{-1}$ to $+50$ km s$^{-1}$, are about as expected near quadrature for the two binaries.}
\label{fig:HIRES}
\end{center}
\end{figure}

The cross correlation between the first of the HIRES spectra and the template from a reference M star is shown in Fig.~\ref{fig:HIRES}.  We clearly detect four significant peaks in the CCF which we identify as belonging to the four M stars in the quadruple star system.  

\subsection{Radial Velocities}
\label{sec:RVcurves}

We cross-correlated the four HIRES and two IGRINS spectra of the red southern image of EPIC 220204960 with high signal-to-noise template spectra of bright, nearby M-dwarfs. For HIRES, we used a spectrum of GL 694, while for IGRINS, we used a spectrum of LHS 533. We placed the cross correlation functions on an absolute velocity frame using the measured absolute RVs of these two template stars from Nidever et al. (2002). 

We summarize in Table \ref{tbl:RVs} all six sets of RV measurements taken at five independent epochs.  In first discussing these measurements we refer to only five sets of measurements since the first of the HIRES spectra is nearly simultaneous in time with the second of the IGRINS spectra.  Thus, in all there are $5 ~{\rm spectra}\,\times 4$ CCF peaks that each must be identified with a particular star in one of the two binaries.  To accomplish this, we chose two peaks from each CCF to represent the stars in binary A, with its known orbital period, temporarily ignoring the other two peaks in the first pass.  We then fit simple circular orbits to $(4 \times 3)^5/2 = 124,416$ distinct combinations of choices of stars with CCF peaks\footnote{The naming convention in the first CCF is a matter of definition, hence the 1/2 factor.}.  Once the CCF-peak to star assignments have been made that work best for binary A, there are only 16 independent combinations remaining to try for binary B.

Each binary fit utilized four free parameters: the two stellar $K$-velocities, $K_1$ and $K_2$, the binary's $\gamma$-velocity, and a linear trend, $\dot \gamma$, to represent possibly detectable acceleration of the binary in its outer orbit.  Only a few such combinations of stellar ID and CCF peak yielded decent $\chi^2$ values and physically sensible results for the binary being fitted, but where the remaining (i.e., unused) CCF peaks could be reasonably fit to the stars in the other binary.  We selected one choice of stellar IDs with CCF peaks that yielded the best fit for {\em both} binaries.  That particular set of RVs matched with stellar components is summarized in Table \ref{tbl:RVs}. 

\begin{figure}
\begin{center}
\includegraphics[width=0.49 \textwidth]{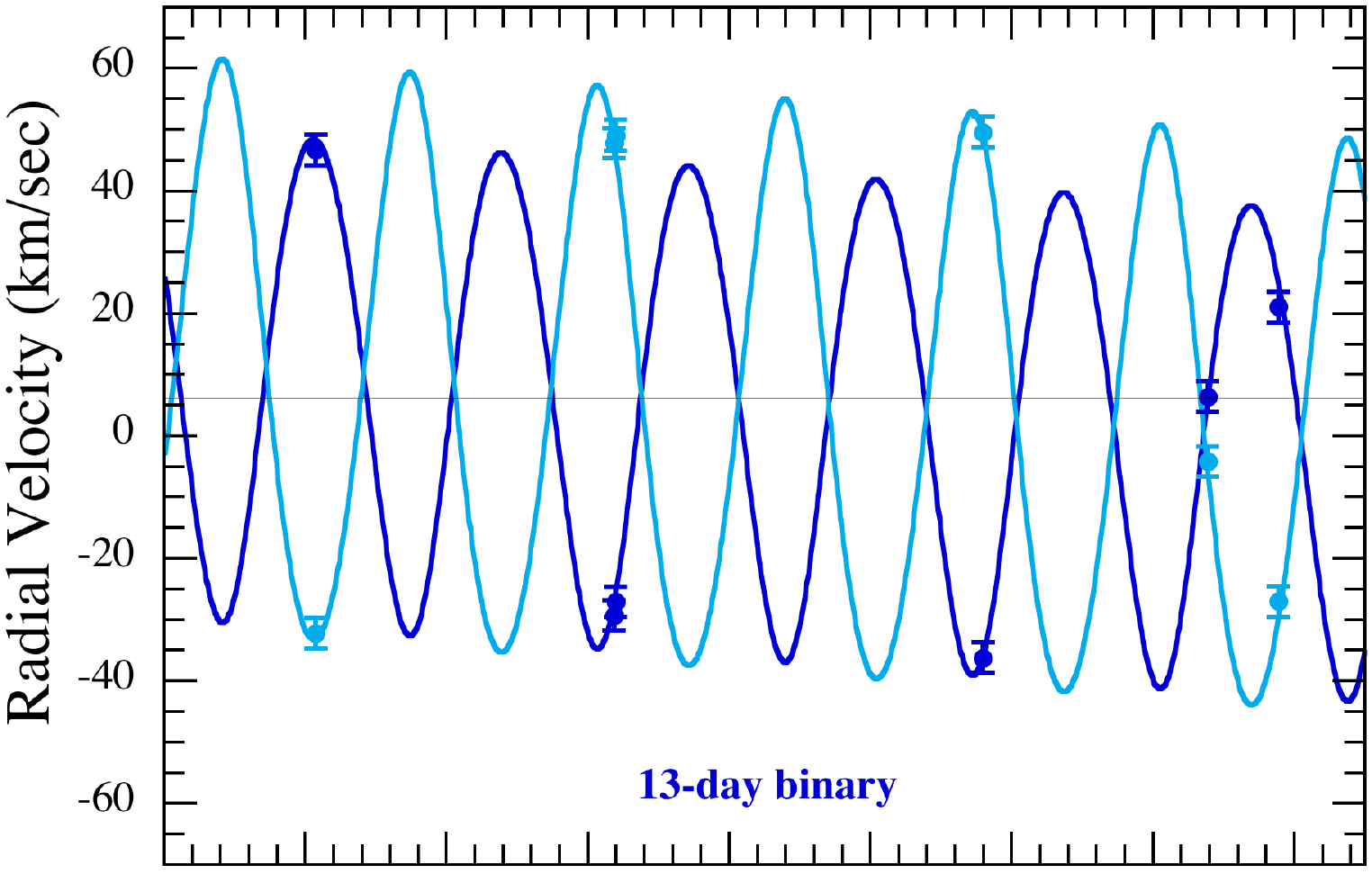} 
\includegraphics[width=0.49 \textwidth]{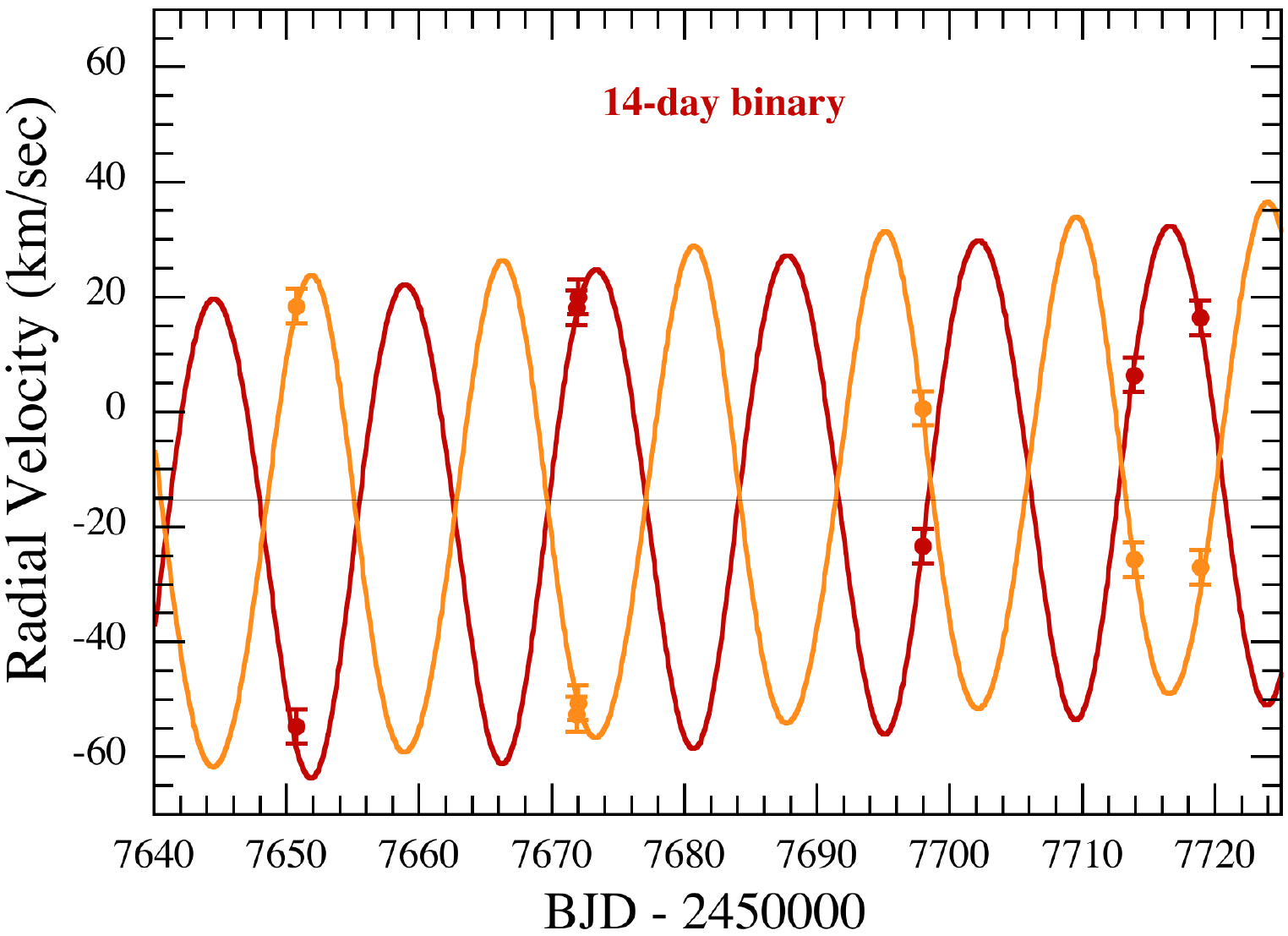}
\caption{Radial velocity measurements from two IGRINS and four HIRES spectra. The second of the IGRINS spectra has nearly the same epoch as the HIRES spectrum.  Top panel is for the 13-d A binary and bottom panel for the 14-d B binary.  The solid curves are the best fits with only the K velocity of each star, the $\gamma$ velocity (black horizontal line), and $\dot \gamma$ as free parameters for each binary, while $\omega$, $\tau$, and $e$ are taken from Table \ref{tbl:simlightcurve}.}
\label{fig:RVs}
\end{center}
\end{figure}

Once the identification of CCF peaks with individual stars has been uniquely made, there are then 10 RV points that are associated with each binary (see Table \ref{tbl:RVs} and Fig.~\ref{fig:RVs}).  In principle, we should then fit these curves with 7 free parameters: $K_1$, $K_2$, $\gamma$, $\dot \gamma$, $\omega$, $\tau$, and $e$, where $\tau$ is the time of periastron passage and, again, $\dot \gamma$ (assumed constant) represents the binary's acceleration in its outer orbit.  In practice, however, we have found that the RV points are neither numerous enough nor sufficiently accurate to derive values for $\omega$ or $e$ that are nearly as good as we are able to derive from the lightcurve analysis (see Sects.~\ref{sec:K2}, \ref{sec:MCMC}, and \ref{sec:lcfactory}).  To a lesser extent, the same is also true of $\tau$.  

We therefore restricted our fits of the RV data points to the four parameters: $K_1$, $K_2$, $\gamma$, and $\dot \gamma$ while fixing $\omega$, $\tau$, and $e$ at the values given in Tables \ref{tbl:MCMC} and \ref{tbl:simlightcurve}.  The fits were carried out with an MCMC routine that is described in more detail in Sect.~\ref{sec:MCMC}.  The results of the fits are shown in Fig.~\ref{fig:RVs} and Table \ref{tbl:RVs}.  The plotted error bars in Fig.~\ref{fig:RVs} are just the empirical rms scatter of the data points about the model curve because we have no other independent way of assessing them.  Note the linear trend ($\dot \gamma$) for both binaries, but of opposite signs, in Fig.~\ref{fig:RVs}.

In addition to the K velocities and uncertainties given in Table \ref{tbl:RVs}, we also list the four constituent stellar masses that we infer from the $K$-velocities.  All four stars seem quite consistent with $\sim$0.4 $M_\odot$ late-K or early-M stars.  We later compare these stellar masses with those found from our analysis of the photometric lightcurves.  The results are in reasonably good agreement and have comparable uncertainties.

The $\gamma$-velocities of the two binaries are found to be: $\gamma_A \simeq +6$ km s$^{-1}$ and $\gamma_B \simeq -14$ km s$^{-1}$.  We can use these two values to compute the `effective' $\gamma$ of the CM of the quadruple system from $\gamma_{\rm quad} \simeq (\gamma_A+\gamma_B)/2 \simeq -4$ km s$^{-1}$.  Since this agrees very well with the $\gamma$ velocity of star `B-N' (see Table \ref{tbl:BN}) we take that as an indication that the two stellar images are part of a physically bound group of five stars.  Finally, with regard to the $\gamma$-velocities, we can also use them to estimate the orbital speed of the two binaries around their common center of mass.  A rough estimate of the instantaneous projected (i.e., radial) speed of each binary in its orbit can be found from $K_{\rm quad} \simeq (\gamma_A-\gamma_B)/2 \simeq 10$ km s$^{-1}$

\begin{figure}
\begin{center}
\includegraphics[width=0.49 \textwidth]{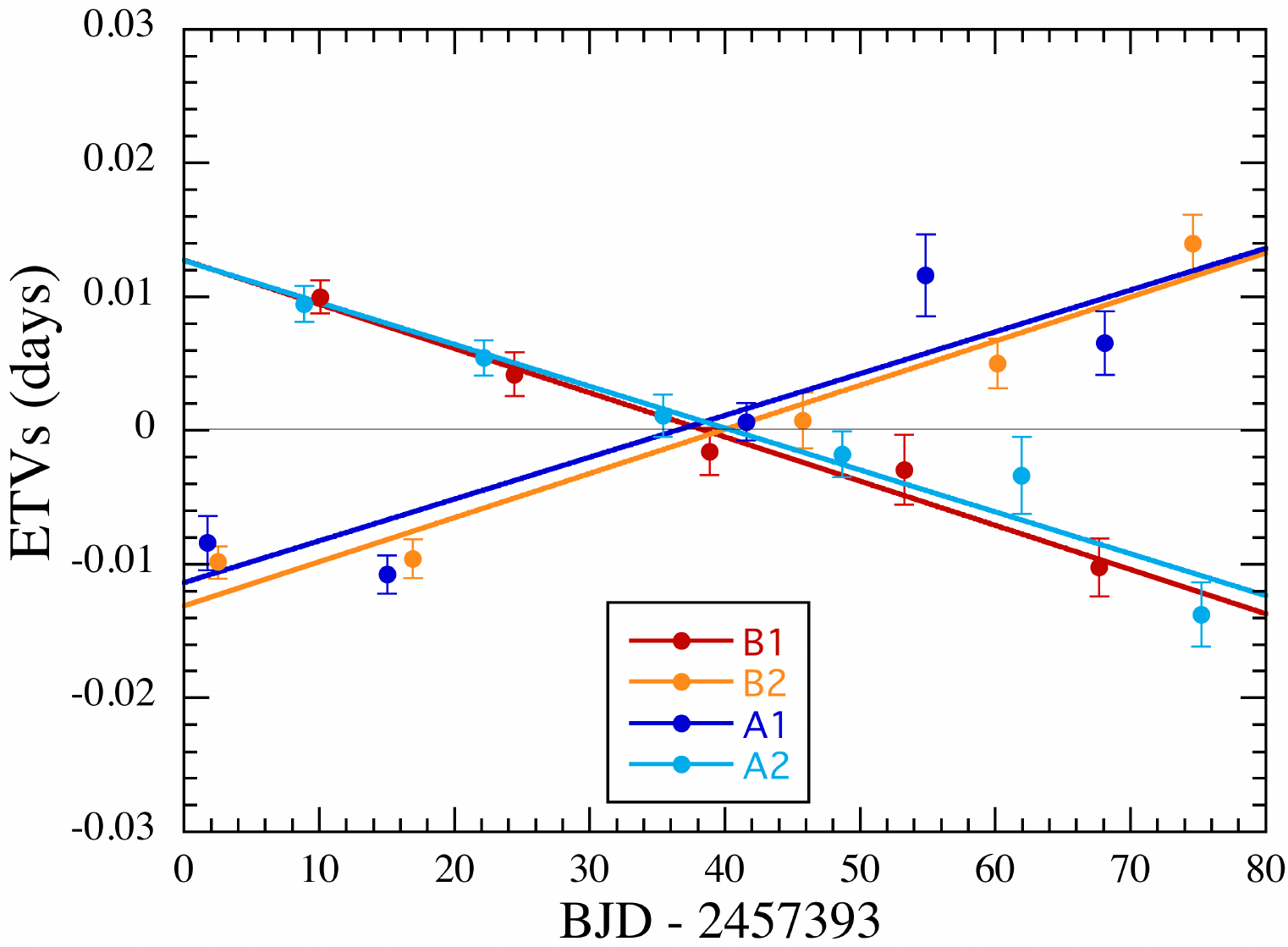}
\caption{Eclipse timing variations in the arrival times of all four eclipses in the two binaries that comprise EPIC 220204960.  In each case the mean orbital period for each binary has been used to produce the ETV curves.  Note the strong divergence of the ETV curves for the primary and secondary eclipses of both the A and B binaries. See Table \ref{tbl:ETVs} for a summary of periods and ETVs.}
\label{fig:ETVs}
\end{center}
\end{figure}   

\begin{figure*}
\begin{center}
\includegraphics[width=0.48 \textwidth]{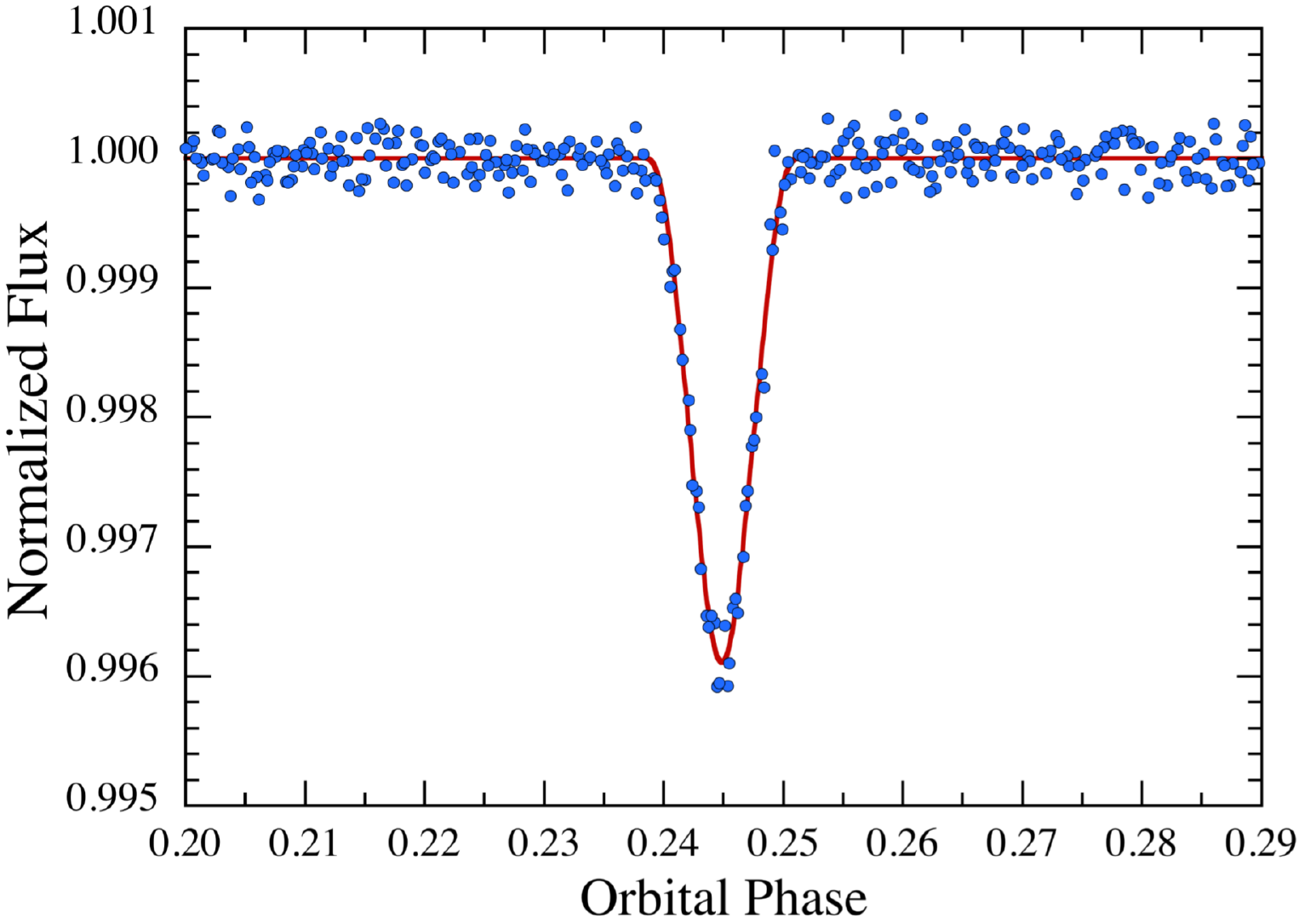} \hglue0.1cm
\includegraphics[width=0.48 \textwidth]{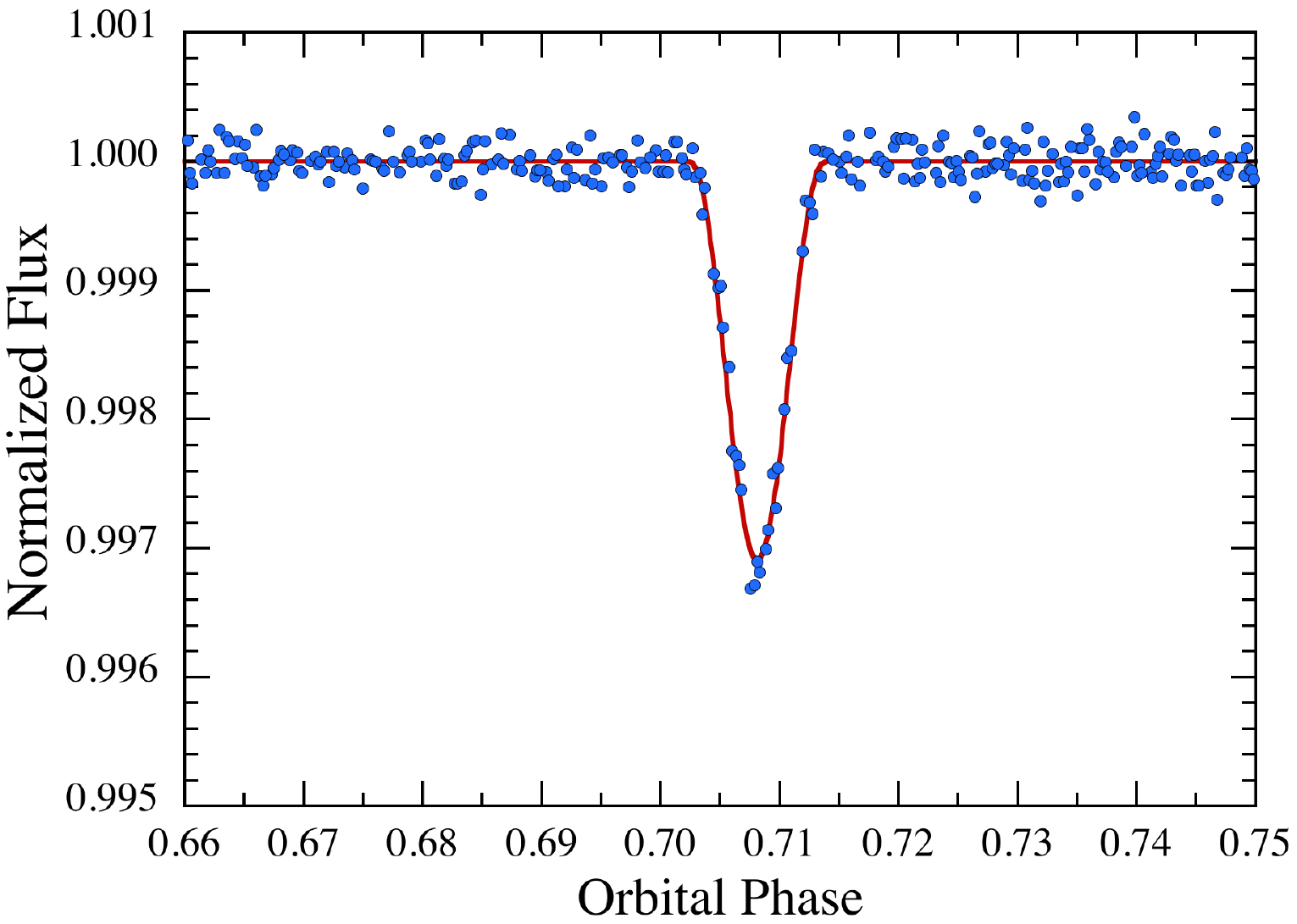} \vglue0.1cm
\includegraphics[width=0.48 \textwidth]{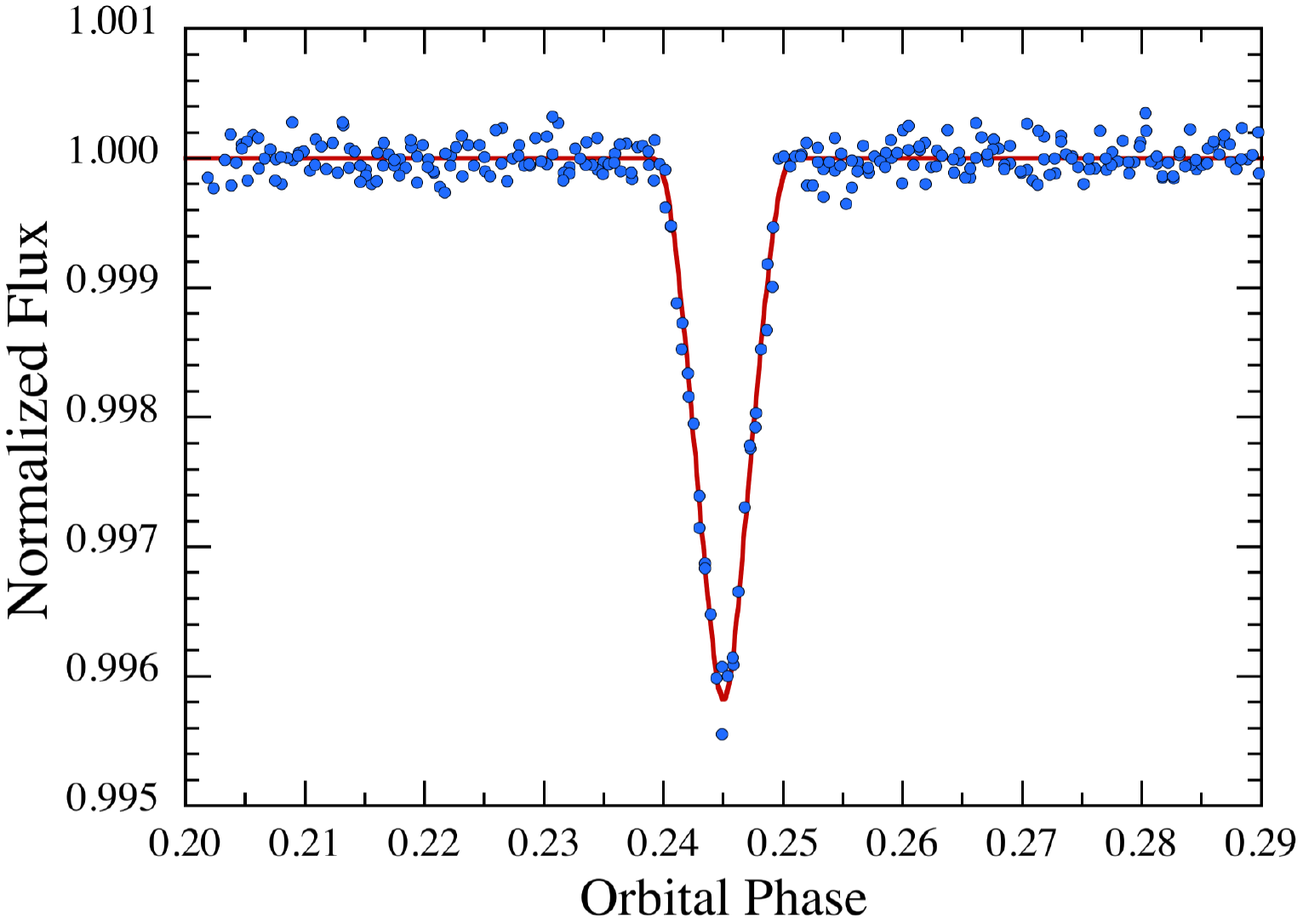} \hglue0.1cm
\includegraphics[width=0.48 \textwidth]{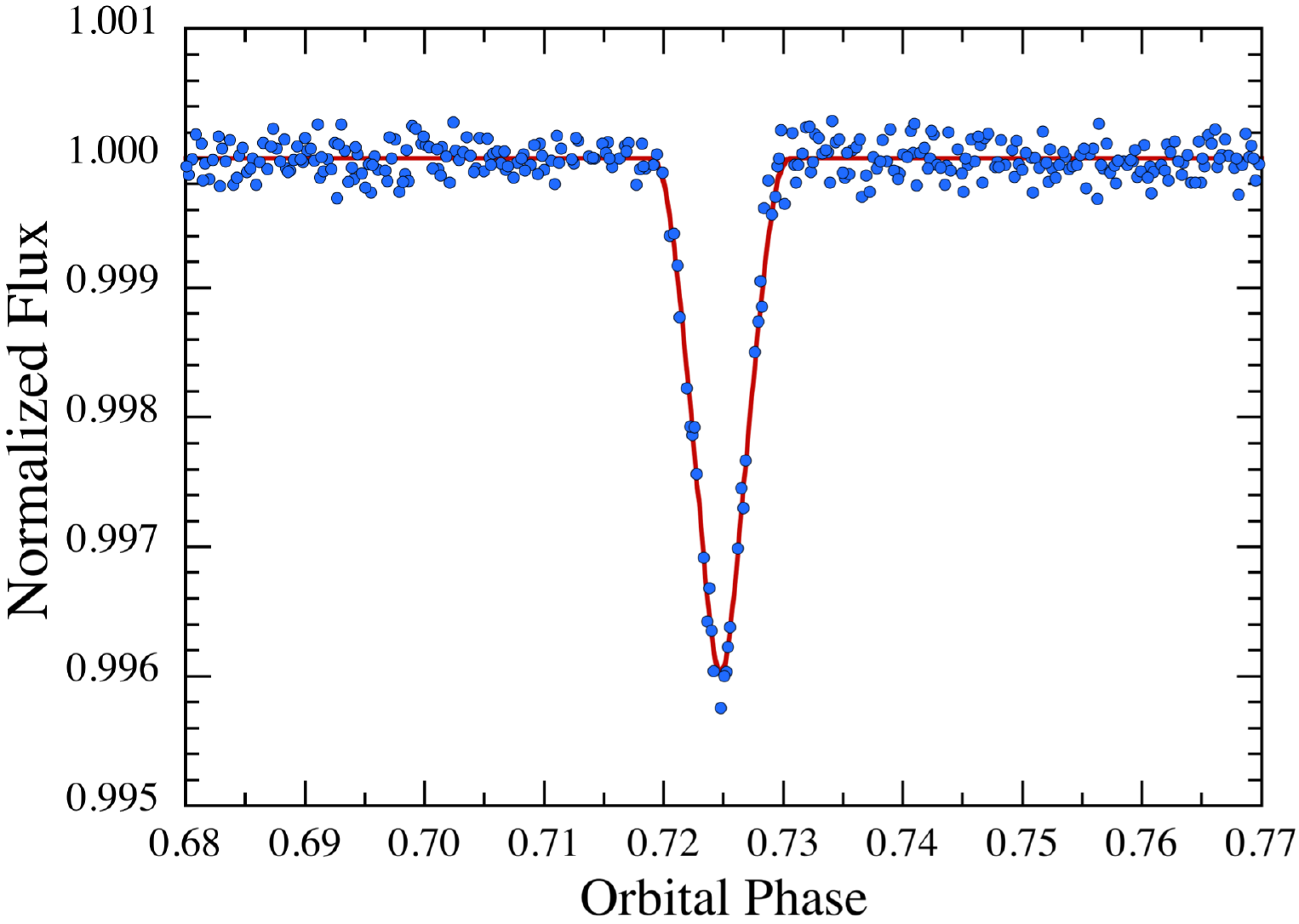}
\caption{K2 eclipse profiles for the primary and secondary eclipses in both the A binary (top panels; $P_{\rm orb} = 13.27$ d) and the B binary (bottom panels; $P_{\rm orb} = 14.41$ d).  Each profile contains data from $\sim$5 eclipses.  Orbital phase zero is arbitrary, but correctly gives the relative phases of the primary and secondary.  The absolute times of the primary eclipse (defined as those in the left panels) are given in Table \ref{tbl:ETVs}.  The red curve is a best fitting model which includes 6 independent parameters for each binary system (see Sect.~\ref{sec:MCMC}).}
\label{fig:eclipses}
\end{center}
\end{figure*}  

\begin{figure}
\begin{center}
\includegraphics[width=0.45 \textwidth]{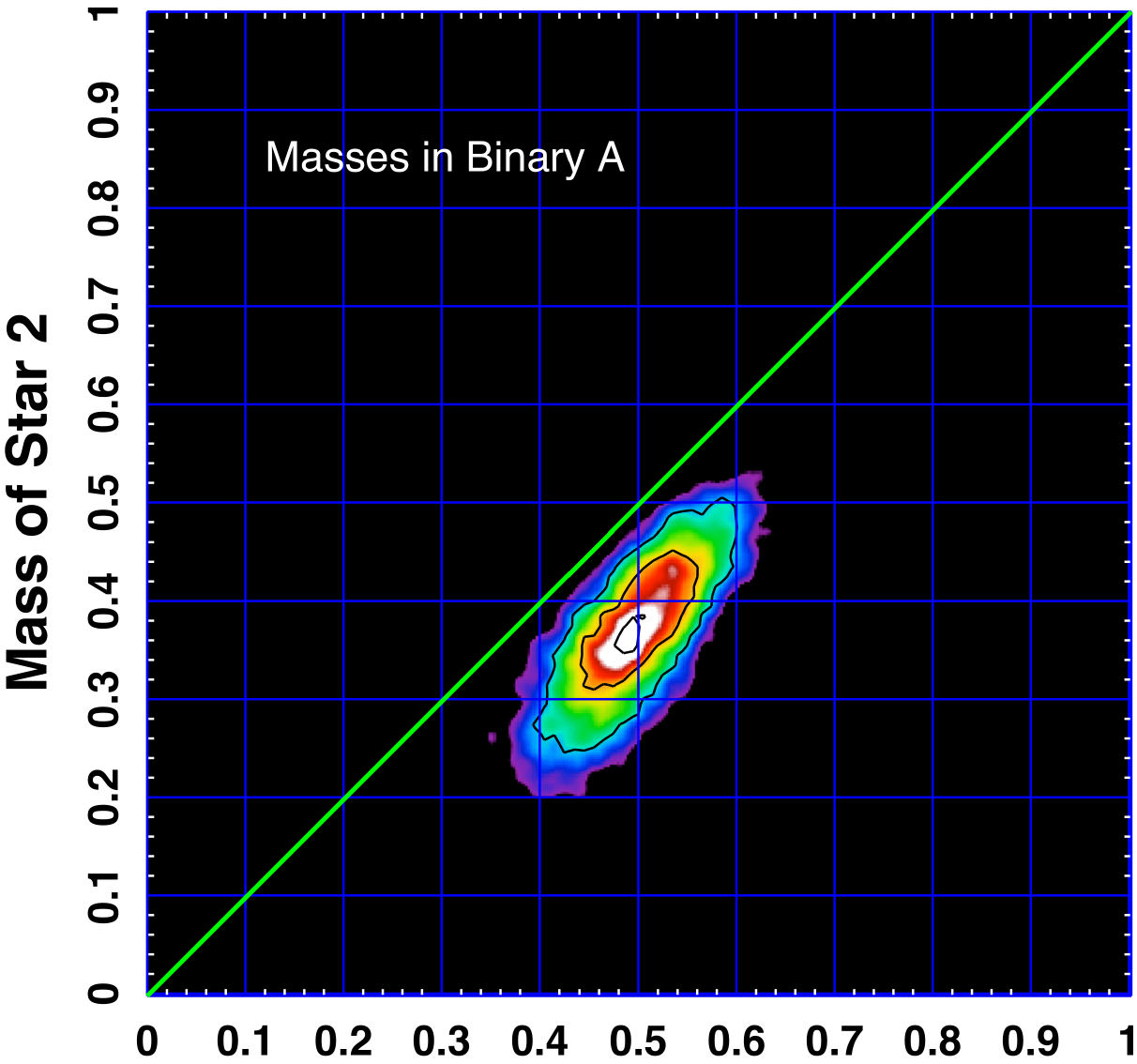}
\includegraphics[width=0.45 \textwidth]{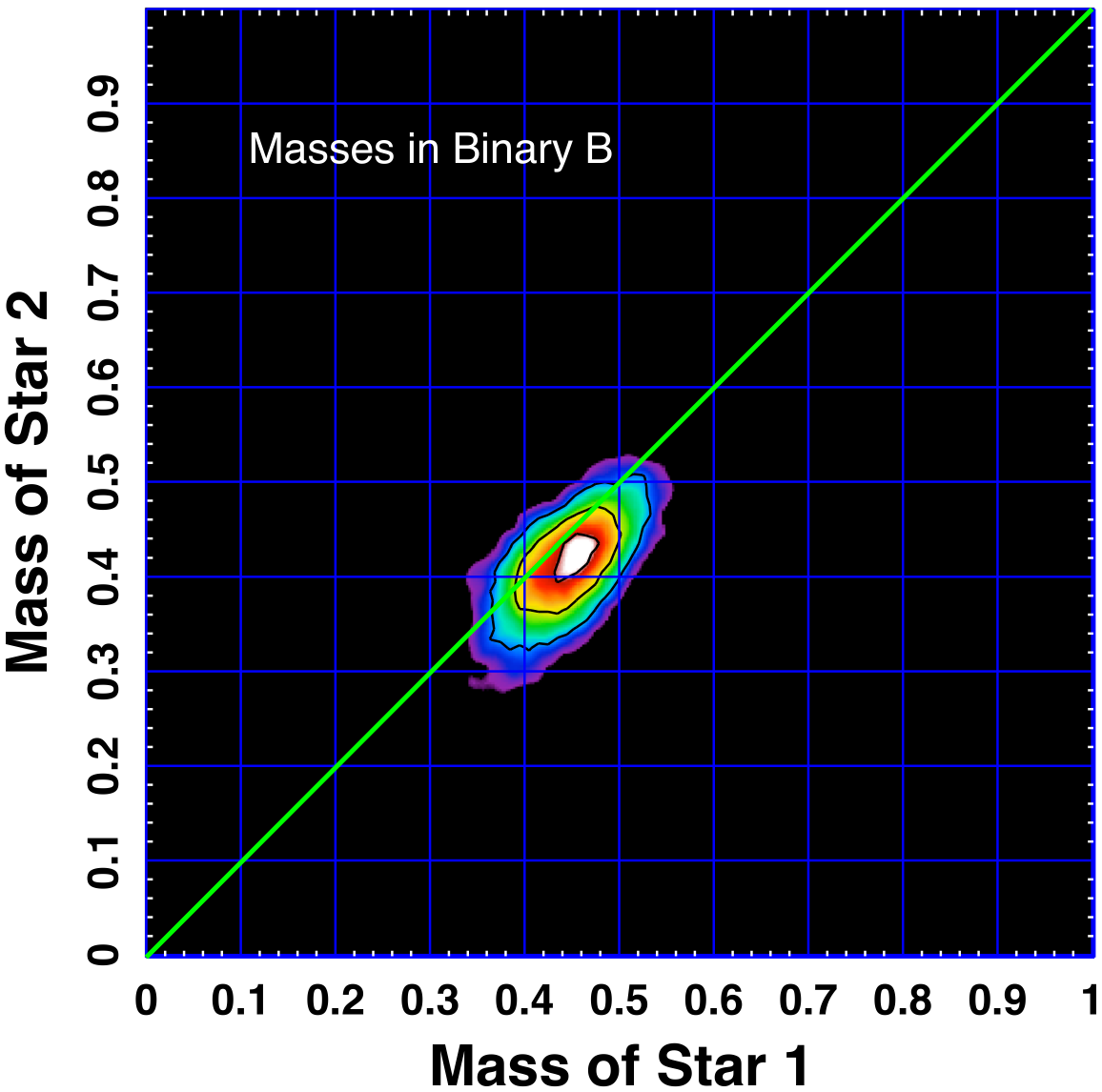}
\caption{Correlation plot between $M_1$ and $M_2$ for both binaries taken from the output of the MCMC fit to the eclipses.}
\label{fig:M1vsM2}
\end{center}
\end{figure}  

\begin{table*}
\centering
\caption{Results from Radial Velocity Study}
\begin{tabular}{lccccc}
\hline
\hline
& Star A-1 & Star A-2 & Star B-1 & Star B-2 &  \\
\hline
Radial Velocity Measurements$^a$: & \\
\hline
BJD-2450000 & & & & & Spectr. \\
\hline
7650.7427  &  +46.7  & $-32.2$  & +18.4   &  $-54.7$  & IGRINS \\
7671.8570  & $-29.3$ &  +47.8  &  $-52.6$  &   +18.1  & IGRINS \\
7671.9812  & $-27.1$  &  +49.1  & $-50.6$ & +20.0  & HIRES \\
7697.9627 &  $-36.3$  &  +49.6  & +0.6  &  $-23.3 $ & HIRES \\
7713.8823 & +6.4  & $-4.2$ & $-25.7$ & $+6.4$ & HIRES \\
7718.8820 & +21.0  & $-27.0$ & $-27.0$ & $+16.4$ & HIRES \\
\hline
Orbit Fits$^b$: & \\
\hline
K [km s$^{-1}$] & $39.5 \pm 2.0$ & $46.5 \pm 2.0$ & $41.6 \pm 2.5$ & $42.8 \pm 2.5$ & \\
$\gamma$$^c$ [km s$^{-1}$]             & \multicolumn{2}{c}{$+6.0 \pm 0.8$}         & \multicolumn{2}{c}{$-13.7 \pm 1.0$} & \\
$\dot \gamma$$^d$ [cm s$^{-2}$]             & \multicolumn{2}{c}{$-0.16 \pm 0.03$}         & \multicolumn{2}{c}{$+0.15 \pm 0.04$} & \\
$\gamma_{\rm quad}$$^e$  [km s$^{-1}$] & \multicolumn{4}{c}{$-3.8 \pm 1.3$}         &  \\
$K_{\rm quad}$$^f$  [km s$^{-1}$]  & \multicolumn{4}{c}{$~9.9 \pm 1.3 $}         &  \\
\hline
Constituent Stellar Masses: & \\
\hline
mass [$M_\odot$]  & $0.47 \pm 0.05$ & $0.40 \pm 0.05$ & $0.45 \pm 0.06$ & $0.44 \pm 0.06$ & \\
\hline
\end{tabular}
\label{tbl:RVs}

{\bf Notes.} (a) Uncertainties are difficult to estimate.  Empirically, we found that error bars on the individual RV values of $\sim$ 3 km s$^{-1}$ yielded good $\chi^2$ values. For a description of how we assigned specific peaks in the cross-correlation to specific stars see text.  (b) The orbit fits for each binary involved four free parameters: $K_1$, $K_2$, $\gamma$, and $\dot \gamma$. The orbital period, eccentricity, and argument and time of periastron were taken from the light-curve analysis (see Table \ref{tbl:MCMC}). (c) Center of mass velocity of each binary. (d) Acceleration of the center of mass of each binary. (e) $\gamma_{\rm quad}$ is the radial velocity of the CM of the entire quadruple system.  This assumes that  masses of the two binaries are approximately equal. (f) $K_{\rm quad}$ is the projected radial speed of either binary in its orbit around the quadruple system.  This also assumes that the masses of the two binaries are the same.

\end{table*}

\section{Eclipse Timing Variations}
\label{sec:ETVs}

In order to analyze the lightcurves, we first folded the data for each binary about the best-determined orbital period.  We quickly discovered, however, that regardless of what fold period we used, one eclipse or the other was misshapen or partially filled in.  This was true for both binaries.  In order to understand the cause, we then fit each of the 20 observed eclipses (approximately 5 each for the primary and secondary eclipses of both binaries), to find accurate arrival times.  

To find the arrival times we fit each eclipse with the following non-physical, but symmetric function (i.e., hyperbolic secant; Rappaport et al.~2014), that has a shape sufficiently close to the eclipse profile, $f(t)$, to allow for a precise measurement of the eclipse center:
\begin{equation}
f(t) \simeq B + 2D\left[\exp[(t-t_0)/w]+\exp[-(t-t_0)/w]\right]^{-1}
\end{equation}
The four free parameters are: $B$, the out-of-eclipse background, $D$, the eclipse depth, $t_0$ the time of the center of the eclipse, and $w$ a characteristic width of the eclipse.

After subtracting off the expected times of eclipse using the mean orbital periods of $P_A = 13.2735$ d and $P_B = 14.4158$ d, we find the eclipse timing variations (hereafter `ETVs') shown in Fig.~\ref{fig:ETVs}.  We were surprised to find that the ETV curves for the primary and secondary eclipses, for both binaries, `diverge' so clearly and by such a large amount over the course of only 80 days.  For both binaries, the divergence in the ETV times amounts to plus and minus 0.025 days for the primary and secondary eclipses, respectively. In terms of slopes to the ETV curves, these correspond to plus and minus $\sim$0.00032 days/day for both binaries, where the plus and minus signs are for the primary and secondary eclipses.  Finally, we can determine an apparent `local' (in time) period for each eclipse.  These are:  13.26913 d, 13.27789 d, 14.41130 d, and 14.42024 d.  These delays, slopes, and apparent periods are summarized in Table \ref{tbl:ETVs}.

Finally, we use these four periods to fold the data, one for each eclipse, in order to produce the eclipse profiles that we use to fit for the orbital parameters.  For these folds we use an epoch near the mid point of the 80-day K2 observations.  Because the primary eclipse profile is produced using a slightly different period from that of the secondary eclipse, the relative phasing between the two eclipses is only well defined at the center time of the K2 observations. However the phase drifts of one eclipse with respect to the other over this time period amount to only $\sim$0.0015 cycles, and thus they do not significantly affect our ability to determine quantities such as eclipse spacing (related to $e \cos \omega$) or the eclipse profiles.  In fact, the meaning of the divergence in the ETVs is precisely the fact that $\omega$ is changing by a small, but measurable amount over the course of the 80-day observation interval.

\begin{table}
\centering
\caption{ETV Divergences and Apparent Orbital Periods$^a$}
\begin{tabular}{lcccc}
\hline
\hline
Parameter &
Star A-1 & 
Star A-2 & 
Star B-1 & 
Star B-2
 \\
\hline
ETV  [d] &  +0.024 & $-0.024$ & $-0.024$ &+0.024 \\ 
ETV slope [d/d] & 0.00033(4) & $-0.00033$ (3) & $-0.00031(4)$ & 0.00031(9) \\ 
$\Delta P_{\rm orb}$ [d] & 0.0044 & $-0.0044$ & $-0.0045$  & 0.0045 \\
Apparent $P_{\rm orb}$ [d] & 13.26913 & 13.27789 & 14.41130 & 14.42024 \\
Epochs [BJD]$^b$ & 7401.864 & 7394.718 & 7403.021 & 7395.497 \\
\hline
\end{tabular}
\label{tbl:ETVs}

{\bf Notes.} (a) Derived from the 20 total eclipses of binaries A and B.  `ETV' refers to the total eclipse timing variations over the 80-day K2 observations.  $\Delta P_{\rm orb}$ refers to the difference between the mean apparent orbital period and that derived independently for the primary and secondary eclipses. (b) The epoch times are actually BJD--2450000.  

\end{table}

\begin{table*}
\centering
\caption{Properties of the Quadruple Stars}
\begin{tabular}{lcccc}
\hline
\hline
Parameter &
\multicolumn{2}{c}{Binary A} & \multicolumn{2}{c}{Binary B}  \\
\hline
$P_{\rm orb}$$^a$ [days]             & \multicolumn{2}{c}{$13.2735 \pm 0.0044$}         & \multicolumn{2}{c}{$14.4158 \pm 0.0045$} \\
semimajor axis$^b$ [$R_\odot$]  &  \multicolumn{2}{c}{$22.8 \pm 1.3$}                     & \multicolumn{2}{c}{$23.7 \pm 0.8$} \\  
inclination$^b$ [deg]                     & \multicolumn{2}{c}{$89.5^{+0.4}_{-0.4}$}            & \multicolumn{2}{c}{$89.7 \pm 0.3$} \\
$e \cos \omega$$^a$                    &  \multicolumn{2}{c}{$0.0577 \pm 0.0001$}            & \multicolumn{2}{c}{$0.0318 \pm 0.0001$} \\  
$e$$^b$                                        &  \multicolumn{2}{c}{$0.061^{+0.017}_{-0.003}$}     & \multicolumn{2}{c}{$0.033^{+0.007}_{-0.002}$} \\ 
$\omega$$^b$ [deg]                     &  \multicolumn{2}{c}{$208^{+20}_{-36}$}                             & \multicolumn{2}{c}{$192^{+32}_{-26}$} \\
$t_{\rm prim~eclipse}$$^a$ [BJD]      &  \multicolumn{2}{c}{$2457401.864 \pm 0.003$}            & \multicolumn{2}{c}{$2457403.021 \pm 0.003$} \\
3rd-light factor$^c$                        & \multicolumn{2}{c}{$90^{+23}_{-25}$}                     & \multicolumn{2}{c}{$97^{+11}_{-19}$} \\
\hline
\hline
individual stars & A1 & A2 & B1 & B2 \\
\hline
mass$^b$ [$M_\odot$] & $0.49^{+0.06}_{-0.07}$ & $0.38^{+0.07}_{-0.09}$          & $0.45^{+0.05}_{-0.06}$ & $0.42^{+0.05}_{-0.06}$ \\
radius$^b$ [$R_\odot$] & $0.45^{+0.05}_{-0.06}$ & $0.35^{+0.06}_{-0.07}$        & $0.41^{+0.04}_{-0.05}$ & $0.39^{+0.04}_{-0.05}$ \\
$T_{\rm eff}$$^b$ [K] & $3600^{+110}_{-80}$ & $3460^{+70}_{-60}$                                       & $3540^{+75}_{-55}$ & $3500^{+60}_{-45}$ \\
luminosity$^b$ [$L_\odot$] & $0.031^{+0.013}_{-0.010}$ & $0.016^{+0.008}_{-0.007}$         & $0.023^{+0.008}_{-0.007}$ & $0.020^{+0.006}_{-0.006}$ \\
$\log \, g$$^b$ [cgs] & $4.82^{+0.06}_{-0.05}$ & $4.92^{+0.08}_{-0.06}$                & $4.86^{+0.05}_{-0.04}$ & $4.89^{+0.05}_{-0.04}$ \\
\hline
\end{tabular}
\label{tbl:MCMC}

{\bf Notes.} (a) Based on the K2 photometry.  (b) Derived from an analysis of the K2 photometric lightcurve (see Sect.~\ref{sec:MCMC}) and the uncertainties are 90\% confidence limits. This analysis utilized the analytic fitting formulae for $R(m)$ and $T_{\rm eff}(m)$ given in equations (\ref{eqn:RM_rel1}) and (\ref{eqn:TM_rel1}).  When we modify the $R(m)$ relation slightly to account for the somewhat larger radii measured for a number of stars in close binaries (see App.~\ref{app:RTLM} and Fig.~\ref{fig:RM}), the cited masses would {\rm decrease} by $\sim$$0.03-0.04 \, M_\odot$.  (c) From photometric measurements of the `B-N' and `R-S' flux ratio.

\end{table*}

The results of folding the data about four different periods leads to the four profiles shown in Fig.~\ref{fig:eclipses}.  Note how similar all four eclipses look in terms of width, shape, and depth. Only the eclipse depth for the secondary star in binary A is perceptibly more shallow than the other three.  In spite of the fact that only a small portion of the lightcurve is shown around each eclipse, the orbital phases of one eclipse with respect to the other, shown on the x axes are correct, at least for the mid-time of the 80-day observation.

\section{Physically Based Fits to Lightcurves}
\label{sec:MCMC}

In this section we fit the lightcurves shown in Fig.~\ref{fig:eclipses} to extract as many of the system parameters as can be constrained by the eclipse depths, shapes, and relative phasing.  We do not attempt to fit the out-of-eclipse regions of the lightcurves for effects such as ellipsoidal light variations (`ELVs'), Doppler boosting, or illumination effects (see, e.g., van Kerkwijk et al.~2010; Carter et al.~2011).  The reasons for this are twofold.  First, with orbital periods as long as 13-14 days, such effects are quite small, i.e., at the $\sim$ten parts per million level (by comparison with the eclipses which are typically 4000 ppm), and these are further seriously diluted by the light from the `B-N' image.  Second, the fidelity of the K2 photometry at these low frequencies, i.e., on timescales of $\gtrsim 10$ days is not to be trusted at these low levels, and in any case they are largely filtered out in the processing of the data.

Because of the very large dilution factor in these eclipses (due to the presence of the `B-N' image in the photometric aperture), the so-called ``3rd light'' (`L3') parameter is in the range 0.985-0.992, as we detail below.  In principle, binary lightcurve emulators such as {\tt Phoebe} (Pr\v{s}a \& Zwitter 2005) can fit for the 3rd light as a free parameter.  In practice, however, we have found that when L3 is so large, and two binary lightcurves are combined photometrically, {\tt Phoebe} is not able to find accurate values for either L3 or the remainder of the binary parameters.  Thus, we adopt a more physically motivated approach to fitting the lightcurves which uses supplemental information to ensure that the L3 parameter is meaningful.  

The approach we utilize to fit the eclipses is closely related to the one presented by Rappaport et al.~(2016) in the study of the quadruple system in EPIC 212651213.  However, it is sufficiently different that we outline our procedure here.

In brief, the goal is to use the information in the two eclipses for each binary, including their orbital phase separation, to fit for 6 free parameters: the two masses, the argument of periastron, the inclination angle, time of periastron passage, and third light.  (The eccentricity is found from the choice of $\omega$ and the already determined value of $e \cos \omega$ -- see Sect.~\ref{sec:K2}.)  We do this under the assumption that all the stars are sufficiently low in mass (i.e., $\lesssim 0.5 \, M_\odot$) that they are substantially unevolved at the current epoch.  This then allows us to determine both the stellar radius and luminosity from the mass (and an assumption about the metallicity).  These 6 parameters are adjusted via a Markov Chain Monte Carlo (`MCMC') routine, which uses the Metropolis-Hastings algorithm (see, e.g., Ford 2005; Madhusudhan \& Winn 2009, and references therein; Rappaport et al.~2016) in order to find the best fitting values and their uncertainties.

In somewhat more detail, each step in the MCMC procedure goes as follows.  We first choose a primary and secondary mass from within a uniform prior ranging from $0.2-0.7 \, M_\odot$. The inclination angle, $i$, is chosen from within a uniform prior ranging from 87$^\circ$ to 90$^\circ$, while the argument of periastron, $\omega$, can range over 0 to 2$\pi$.  The dilution factor for either binary is chosen from within the range 60--120 (equivalent to a 3rd light of 0.987--0.992).  Note that because there are two binaries within the `R-S' image, this dilution factor is about twice the ratio of fluxes we find for `B-N'/`R-S'.  Finally, the time of periastron passage, $\tau$, is chosen over a small range based on the fact that for nearly circular orbits $\tau \simeq t_{\rm ecl}  + P_{\rm orb}(\omega/2 \pi-1/4)$ where $t_{\rm ecl}$ is the eclipse time.  

Once the masses have been chosen, we compute the orbital separation from Kepler's 3rd law using the known orbital period.  The stellar radii and effective temperatures are calculated from analytic fitting formulae for low-mass main-sequence stars.  Initially, we utilized the expressions of Tout et al.~(1996) which cover the entire main sequence (0.1 -100 $M_\odot$), but later switched to our own relations derived more explicitly for stars on the lower main sequence.  We later verified that the two sets of fitting formulae actually produce fairly similar results.  Our fitting formulae for $R(m)$ and $T_{\rm eff}(m)$, discussed in Appendix A, are of the form:
\begin{equation}
\log[R(m)]  =  \sum_{n =1}^5 c_n \log^n(m) \\ \label{eqn:RM_rel1}
\end{equation}
\begin{equation}
T_{\rm eff}(m)  =  \frac{b_1 m^{4.5}+b_2 m^{6}+b_3 m^{7}+b_4 m^{7.5}}{ 1 + b_5 m^{4.5} +b_6 m^{6.5} }  ~{\rm K}
\label{eqn:TM_rel1}
\end{equation}
where $m$ is the mass in $M_\odot$, $R$ is in units of $R_\odot$, and the constant coefficients $c_n$ and $b_n$ are given in the Appendix.  

The binary lightcurve is generated for two spherical stars that are limb darkened with a quadratic limb-darkening law using coefficients appropriate for early M stars and taken from Claret \& Bloemen (2011).  As discussed above, no ELVs, illumination, or Doppler boosting effects were computed because the wide orbit and the low frequency behavior of these features would not reveal such effects.  The lightcurve was computed in 2-minute steps, and then convolved with the {\em Kepler} long cadence time of 29.4 minutes.  

After the MCMC parameters have been chosen for a given binary realization, we use the value of the dilution factor in the current MCMC step to scale the model lightcurve accordingly. 

The model lightcurves are then compared to the observed lightcurves with $\chi^2$ as the quantitative measure of agreement. The Metropolis-Hastings jump conditions (see, e.g., Ford 2005) are then used to decide whether a given step will be accepted or not.  If the step is accepted, then that set of parameters is stored as part of the parameter distributions.

After this process has been repeated many times, the probability distributions for the parameters of both stars in the binary under consideration, as well as those for $i$, $\omega$, and $e$ are evaluated.  The best fitting values and their $1 \, \sigma$ uncertainties are listed in Table \ref{tbl:MCMC}.  The best fits to the lightcurves of the two binaries are shown in detail in Fig.~\ref{fig:eclipses}. 

Most of the parameter uncertainties, as determined from the MCMC analysis, are unremarkable, and are given in Table \ref{tbl:MCMC}.  However, in the case of Binary A, the masses of the two stars are significantly different.  In Fig.~\ref{fig:M1vsM2} we show the correlation between $M_1$ and $M_2$ in both binaries.  Note that the region of uncertainty in the $M_1-M_2$ plane for Binary A lies entirely below the $M_1=M_2$ line.  This is related to the fact that for low-mass stars in the mass range $0.3-0.5 \, M_\odot$ the $T_{\rm eff}(M)$ relation is remarkably flat (Baraffe \& Chabrier 1996; Baraffe et al.~1998).  Since the ratio of eclipse depths depends only on the values of $T_{\rm eff}$ for the two stars (assuming circular orbits or where $\omega \simeq 0$), in order to explain the 25\% more shallow eclipse depth for star 2 in Binary A, the fitting code needs to considerably reduce the values of $M_2$ compared to $M_1$.

A rather clear picture emerges of four quite similar M-stars in two impressively alike binaries.  We were gratified that so much information could be extracted from the measurement of only $\sim$10 eclipses for each binary.  In particular, we find impressive agreement with the masses derived independently from the RV measurements (Table \ref{tbl:RVs}), and note that the uncertainties of both determinations are actually quite comparable.

Finally, in regard to the MCMC fits, we have run the code with the dilution factor as a free parameter with a large prior range of values (i.e., $\gtrsim 60$) as well as with a narrow enough range so as to force a match with the observed ratio of the `B-N'/`R-S' fluxes (i.e., dilution = $90 \pm 20$).  The extraction of the basic physical results for the binaries is affected only in an incidental way.

\begin{figure*}
\begin{center}
\includegraphics[width=0.48 \textwidth]{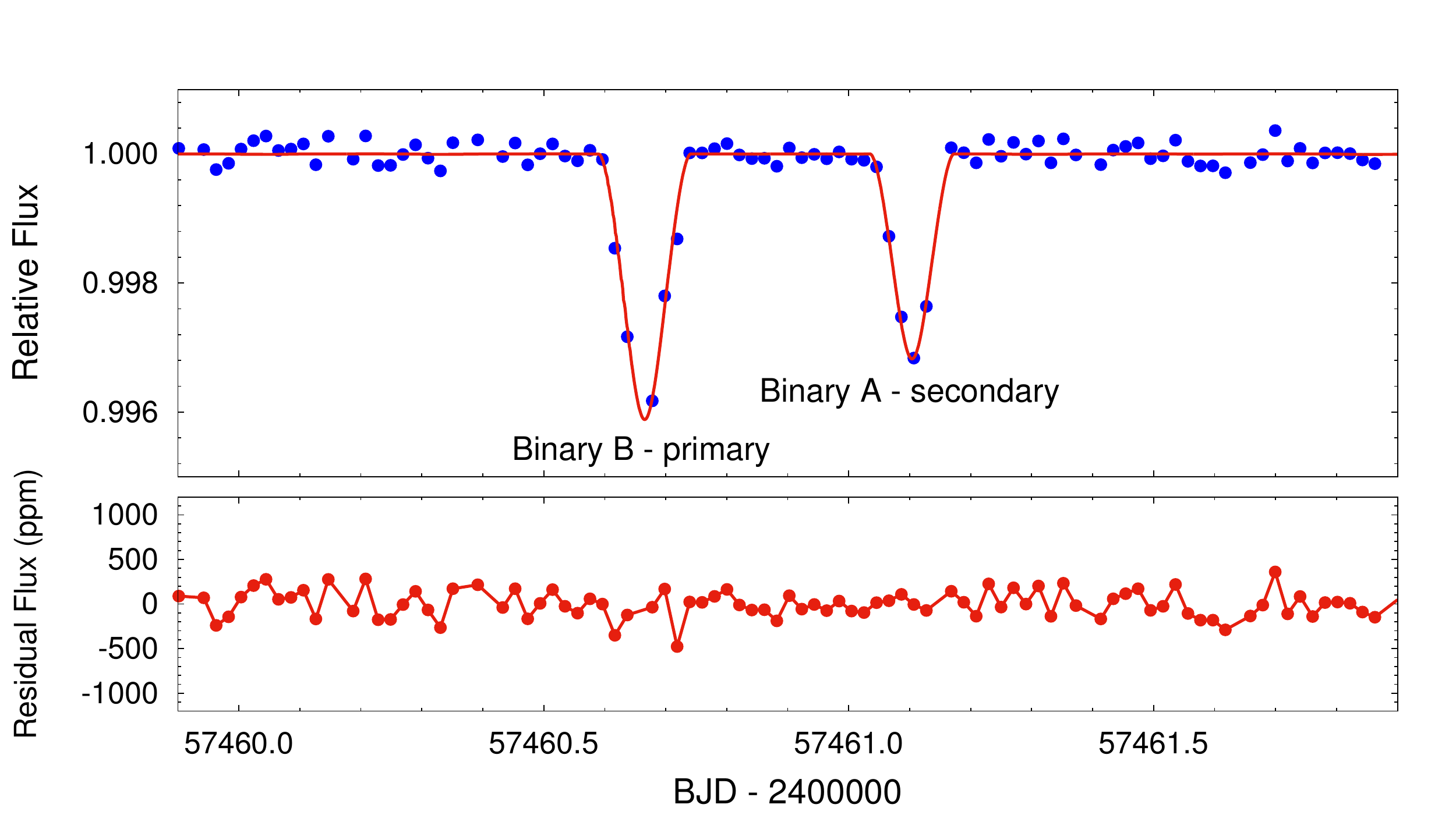} \hglue0.1cm
\includegraphics[width=0.48 \textwidth]{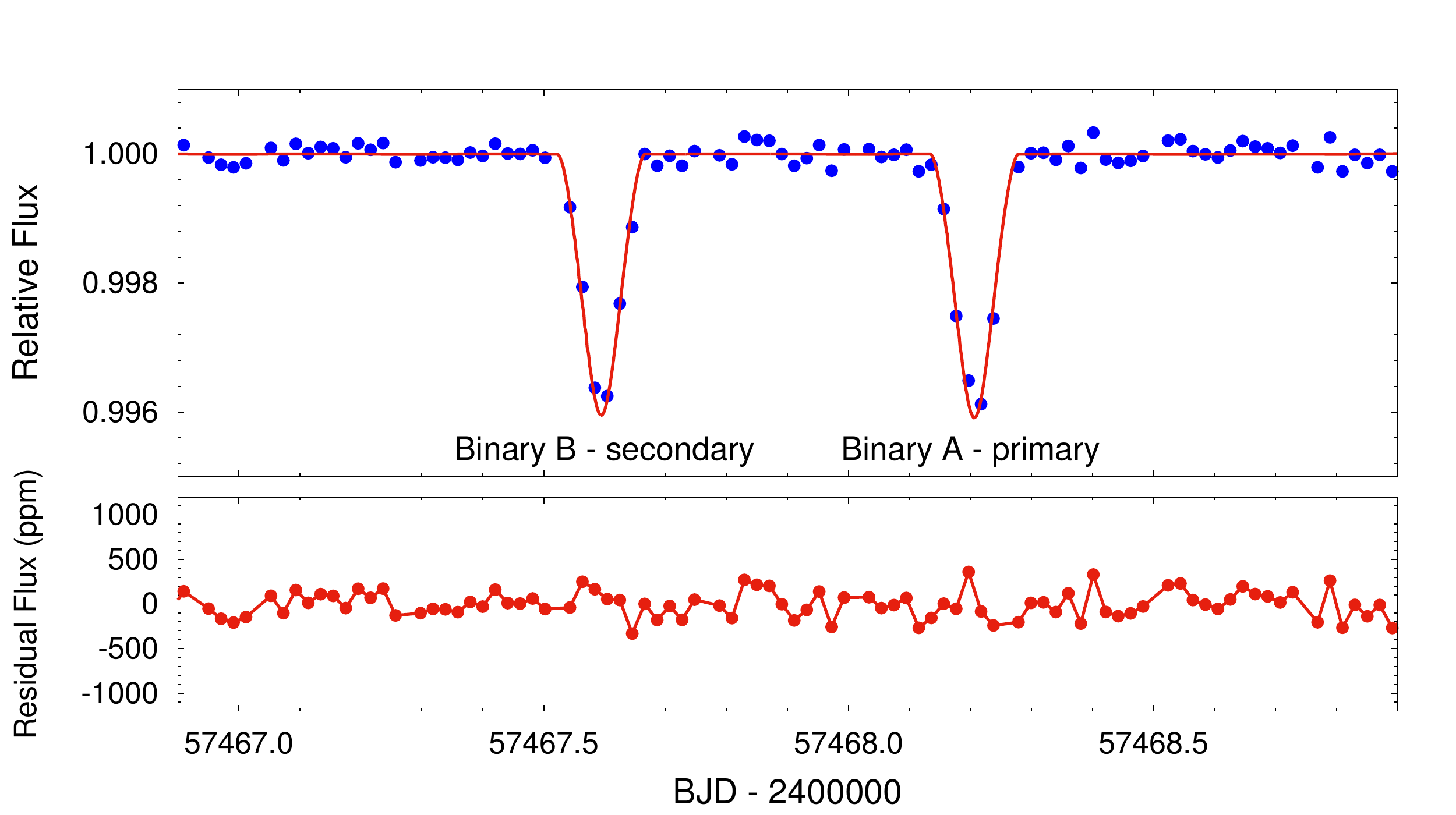} 
\caption{Characteristic portions of the K2 lightcurve together with the synthetic simultaneous solution lightcurve (upper panel), and the residual curve (below).  Note that the two eclipses shown in each panel are for two different binaries.}
\label{fig:simlightcurve}
\end{center}
\end{figure*}  

\section{Simultaneous lightcurve solution}
\label{sec:lcfactory}

In this section we present an approach to simultaneously modeling the lightcurves of two interacting binaries within a single photometric aperture. As we shall see, this approach is quite complementary to the physically-based lightcurve solutions discussed in Sect.~\ref{sec:MCMC}.  For this purpose we modified our Wilson-Devinney- and {\tt Phoebe}-based lightcurve emulator (see~e.g., Wilson~\&~Devinney 1971; Wilson 1979; Wilson 2008; Pr\v{s}a \& Zwitter 2005), {\tt Lightcurvefactory} (Borkovits et al.~2013), to solve both binary lightcurves simultaneously. The practical difficulty of such a simultaneous analysis is that it requires at least twice the number of parameters to be adjusted (or even more) than in a traditional analysis of a single EB lightcurve (in this regard, see the discussion of Caga\v{s}~\&~Pejcha~2012, which to our knowledge is the only prior paper which reports a simultaneous lightcurve analysis of two blended EBs).  However, when either overlapping eclipses are present, or there are large out-of-eclipse variations in the lightcurve(s) which make the simple, phase-folding-based disentanglement (see, e.g., Rappaport et al.~2016) impossible, a simultaneous analysis becomes inevitably important.  In our current situation this is not the case. As was illustrated in the previous sections, the lightcurves of the two EBs can, by chance, be nicely separated. On the other hand, an important coupling remains between the two lightcurves even in this case, namely the flux ratio of the two EBs. If the two EB-lightcurves are solved separately for the two systems it would mean that the value of the `third-light' parameter in each solution would depend on the results of the complete solution for the other EB.\footnote{Strictly speaking, another coupling between the two blended lightcurves comes from both the light-travel time effect and the short timescale gravitational perturbations (see Sect.\,\ref{sec:numerical}) arising from the outer orbit of the two EBs which form a tight quadruple system. These effects, however, cannot be modeled due to the insufficient length of the observed dataset; therefore, we do not take them into account with the only exception being the linear approximation of the dynamically forced apsidal motion.} 

Another reason for carrying out this additional simultaneous lightcurve analysis is to model the rapid, dynamically forced, apsidal motion in both binaries. While the previously applied physically based lightcurve fit (see Sect.~\ref{sec:MCMC}) is found to be highly effective in the quick and accurate determination of the fundamental astrophysical parameters of the binary members, in that method the effect of the apsidal motion was averaged out.  This resulted in larger uncertainties in the other orbital parameters, especially in the arguments of periastron ($\omega_{A,B}$) and in the eccentricities ($e_\mathrm{A,B}$) because of the use of eclipses that were averaged both in their locations and durations.
  
The practical difficulties of the present simultaneous, but otherwise traditional, lightcurve analysis are twofold.  First, the apsidal motion of the binaries should be modeled over the complete 80-day-long K2 lightcurve (Fig.~\ref{fig:rawLC}), i.e., the solution lightcurve should be calculated for all the individual eclipses, instead of calculating the solution only for the four averaged eclipsing lightcurves (Fig.\,\ref{fig:eclipses}).  The other reason lies in the large number of parameters that need to be adjusted. For example, as will be discussed just below, in our case the number of required parameters to be adjusted is about 20. For this reason we made an effort to reduce the number of free parameters and therefore to save computational time. Thus, we took into account five strictly geometrical constraints among some of the parameters. 

\begin{table*}
\centering
\caption{Parameters from the 2EB simultaneous lightcurve solution}
\begin{tabular}{lcccc}
\hline
\hline
Parameter &
\multicolumn{2}{c}{Binary A} & \multicolumn{2}{c}{Binary B}  \\
\hline
$P_{\rm sid}$ [days]             & \multicolumn{2}{c}{$13.2737 \pm 0.0005$}         & \multicolumn{2}{c}{$14.4161 \pm 0.0004$} \\
$P_{\rm anom}$ [days]            & \multicolumn{2}{c}{$13.3491 \pm 0.014$}         & \multicolumn{2}{c}{$14.5122 \pm 0.017$} \\
semimajor axis$^a$  [$R_\odot$]  &  \multicolumn{2}{c}{$ 22.64\pm 0.74$}           & \multicolumn{2}{c}{$ 24.18\pm1.01 $} \\  
$i$ [deg]                & \multicolumn{2}{c}{$89.30 \pm 0.39$}        & \multicolumn{2}{c}{$89.58 \pm 0.45$} \\
$e$                      &  \multicolumn{2}{c}{$0.0636 \pm 0.0016$}     & \multicolumn{2}{c}{$0.0400 \pm 0.0018$} \\  
$\omega_0^b$ [deg]       &  \multicolumn{2}{c}{$201.1 \pm 3.8$}      & \multicolumn{2}{c}{$211.1 \pm 4.4$} \\
$\dot\omega^c$ [deg/yr]  &  \multicolumn{2}{c}{$56.3 \pm 10.7$}      & \multicolumn{2}{c}{$60.8 \pm 11.0$} \\
$\tau^b$ [BJD]           &  \multicolumn{2}{c}{$2457392.341\pm0.005$} & \multicolumn{2}{c}{$2457393.207 \pm 0.003$} \\                  
$t_{\rm prim~eclipse}$ [BJD]&\multicolumn{2}{c}{$2457401.859\pm0.005$}& \multicolumn{2}{c}{$2457403.020 \pm 0.003$} \\
\hline
individual stars & A1 & A2 & B1 & B2 \\
\hline
Relative Quantities: & \\
\hline
mass ratio$^a$ [$q=m_2/m_1$] &\multicolumn{2}{c}{$0.85\pm0.06$} & \multicolumn{2}{c}{$0.97\pm0.08$} \\
fractional radius [$R/a$]& $0.02004\pm0.00185$ & $0.01607\pm0.00169$ & $0.01680\pm0.00132$ & $0.01654\pm0.00131$ \\
fractional luminosity  & $0.00848$ & $0.00405$ & $0.00563$ & $0.00504$ \\
extra light  [$l_5$]   &\multicolumn{4}{c}{$0.977^{+0.005}_{-0.019}$} \\
\hline
Physical Quantities: &  \\ 
\hline
$T_{\rm eff}^d$ [K] & $3564\pm69$ & $3361\pm97$ & $3471\pm112$ & $3421\pm135$ \\
mass$^a$ [$M_\odot$] & $0.47\pm0.05$ & $0.40\pm0.04$  & $0.46\pm0.06$ & $0.44\pm0.06$ \\
radius$^e$ [$R_\odot$] & $0.45\pm0.04$ & $0.36\pm0.04$  & $0.41\pm0.04$ & $0.40\pm0.04$ \\
radius$^f$ [$R_\odot$] & $0.43\pm0.06$ & $0.37\pm0.05$  & $0.42\pm0.08$ & $0.41\pm0.08$ \\ 
luminosity$^g$ [$L_\odot$] & $0.036\pm0.014$ & $0.018\pm0.007$  & $0.026\pm0.010$ & $0.024\pm0.009$ \\
$\log \, g$$^g$ [cgs] & $4.80\pm0.10$ & $4.92\pm0.11$  & $4.88\pm0.10$ & $4.88\pm0.10$ \\
\hline
\end{tabular}
\label{tbl:simlightcurve}

{\bf Notes.} (a) Derived from the RV solution; (b) Derived by the use of geometrical constrains, discussed in the text; (c) Determined for the epoch $T_0=2\,457\,401.8642$; (d) $T_{\rm eff,A1}$ was inferred from Eqn.~(\ref{eqn:TM_rel1}), while the others were calculated from the temperature ratios; 
(e) Stellar radii were derived from the fractional radii and the orbital separation inferred from the RV solution; (f) Stellar radii were derived directly from the $R(m)$ expression in Eqn.\,(\ref{eqn:RM_rel1}) and the masses obtained from the RV solution; (g) Derived quantities using the first set of stellar radii (determined from the $R/a$ values). 
\end{table*}

These constraints are as follows. From the K2 lightcurve we determined the mid-eclipse times of the first primary and the first secondary eclipse for both binaries and also the durations of the first primary eclipses.  We then used these results to constrain the periastron passage times ($\tau_\mathrm{A,B}$), arguments of periastron ($\omega_\mathrm{A,B}$), and the sum of the fractional stellar radii $(R_\mathrm{A1,B1}+R_\mathrm{A2,B2})/a_\mathrm{A,B}$ in the following manner. 
First, for the time offset of a secondary eclipse with respect to the previous primary eclipse we used an extended third-order relation (i.e., taking into account the weak inclination dependence, as well) which, according to  Gim\'enez~\&~Garcia-Pelayo~(1983) is given by:
\begin{equation}
\Delta{T}=0.5P_\mathrm{s}+\frac{P_\mathrm{a}}{\pi}\left[2F_1(e,i) \, e\cos\omega-\frac{1}{3}F_3(e,i) \, e^3\cos3\omega+\mathcal{O}(e^5)\right],
\label{eqn:ecosom}
\end{equation}
where the relation between the sidereal (or, eclipsing) $P_\mathrm{s}$ and anomalistic $P_\mathrm{a}$ periods is
\begin{equation}
P_\mathrm{s}=P_\mathrm{a}\left(1-\frac{\Delta\omega}{2\pi}\right),
\end{equation}
and $\Delta\omega$ stands for the apsidal motion during one revolution of the binary. Furthermore, functions $F_{1,3}$ describe the very weak (practically negligible) inclination and eccentricity dependence of the occurrence times of eclipsing minima (see Eqn.\,20 of Gim\'enez~\&~Garcia-Pelayo, 1983). Solving Eqn.\,(\ref{eqn:ecosom}) which is third order in $e\cos\omega$, the argument of periastron $\omega$ can be determined at each step for the given values of parameters $P_\mathrm{s},e,i,\dot\omega$.

Second, using the relation that at the mid-times of each eclipse the true anomaly takes the value of 
\begin{equation}
\phi =\pm90\degr-\omega+\delta(e,\omega,i),
\label{eqn:f_ecl}
\end{equation} 
and any mid-eclipse times can simply be converted into the actual time of periastron passage ($\tau$) with the use of the Kepler's equation. Note, in our case
\begin{equation}
\delta\approx\pm\frac{e\cos\omega\cos^2i}{\sin^2i\pm e\sin\omega}\ll1
\end{equation}
(see e.g. Borkovits~et~al.~2015, Eqn.\,26) and therefore, negligible.

Finally, from the Taylor expansion of the projected separation of the centres of the stellar disks at the times of the
first and last contact one finds that
\begin{eqnarray}
\left(\frac{R_1+R_2}{a}\right)^2&=&\left(\frac{1-e^2}{1\pm e\sin\omega}\right)^2\cos^2i+\left[\frac{\left(1\pm e\sin\omega\right)^2}{1-e^2}\right. \nonumber \\
&&\left.-\left(1\pm e\sin\omega\frac{1\pm e\sin\omega}{1-e^2}\right)\cos^2i\right]\frac{\pi^2}{P_\mathrm{a}^2}\left(\Delta t\right)^2\nonumber \\
&&+\mathcal{O}\left(\cos^4i\right),
\label{eqn:frac_rad_sum}
\end{eqnarray}
where $\Delta t$ stands for the total duration of the given eclipse and, as above, the upper signs refer to that eclipse which occurs around $\phi+\omega\approx+90\degr$.

With the use of the above relations, the number of parameters to be adjusted was reduced to 14. Eight of them are the orbital parameters $P_\mathrm{A,B}$, $e_\mathrm{A,B}$, $i_\mathrm{A,B}$, including the apsidal advance rates $\dot\omega_\mathrm{A,B}$. Another four star-specific adjusted parameters are the ratios of stellar radii $(R_2/R_1)_\mathrm{A,B}$ and temperatures $(T_2/T_1)_\mathrm{A,B}$ within each binary. Furthermore, the temperature ratio of the two primaries ($T_\mathrm{B1}/T_\mathrm{A1}$) was also adjusted. Finally, we also allowed the extra light in the system to be adjusted, which in this special case should be denoted as $l_5$. On the other hand, we decided not to adjust the mass ratio $q_\mathrm{A,B}$, but rather to fix it at an arbitrary value near unity. This can be justified by the fact that in the case of such widely detached systems (the fractional radii of all four stars were found to be $\leq0.02$) the effect of the tidal forces (having a cubic relation to the fractional radii) on the stellar shapes remains negligible.  Therefore, neither the eclipse geometry, nor the out-of-eclipse region (where ellipsoidal light variations would be found) is influenced by the mass ratios (via tidal distortion), and thus $q_\mathrm{A,B}$ is practically unconstrained photometrically. (Note, the same fact also provides a good justification for the use of the relation in Eqn.\,(\ref{eqn:frac_rad_sum}) which remains valid only insofar as the stellar disks are undistorted.) We also ceased to take into account the relations $R(m)$ and $T_{\rm eff}(m)$ in Eqns.\,(\ref{eqn:RM_rel1}) and (\ref{eqn:TM_rel1}) which had played a key role in the physical lightcurve solution of the previous section. In such a way, in the present analysis, we used only strictly geometrical constraints, and omitted the inclusion of any dimensioned astrophysical quantities. 

Considering other parameters, a quadratic limb-darkening law was applied, for which the coefficients were interpolated from the passband-dependent precomputed tables of the {\tt Phoebe} team\footnote{http://phoebe-project.org/1.0/?q=node/110} (Pr\v sa~et~al.~2011). Note, these tables are based on the results of Castelli~\&~Kurucz (2004). The gravity darkening exponents were set to their traditional values appropriate for such late-type stars ($g=0.32$). The illumination and Doppler-boosting effects were neglected. Even though the calculation of the ellipsoidal light variations is inherent to the code, as mentioned above, they do not play any role.

The results of the simultaneous lightcurve analysis are tabulated in Table\,\ref{tbl:simlightcurve}. Short illustrative sections of the solution lightcurve are also presented in Fig.\,\ref{fig:simlightcurve}. 

Besides the directly adjusted and the geometrically constrained quantities, we can take an additional step and also derive some important physical parameters. The combination of the photometrically determined inclination with the RV solution yields the stellar masses. If the masses are known, the semi-major axes can be determined from Kepler's third law. (We should keep in mind that for a precise result the {\em anomalistic} periods should be used; however, this is of theoretical, but not practical, importance). In the next step, stellar radii can be derived from their dimensionless fractional counterparts, and other quantities (e.g., $\log\,g$'s) can also be calculated. Then, the last free physical parameter, i.e., $T_\mathrm{eff,A1}$ was calculated from the $T_{\rm eff}(m)$ relation given by Eqn.~(\ref{eqn:TM_rel1}).
Once $T_\mathrm{eff,A1}$ is known, the temperatures  of the other three stars can be directly calculated from the direct outputs (i.e. the temperature ratios) of the lightcurve solution.

Note, the existence of the $R(m)$ and $T_{\rm eff}(m)$ relations (Eqns.~\ref{eqn:RM_rel1} and \ref{eqn:TM_rel1}) provides an additional possibility for probing either the astrophysical reliability of our solution, or the validity of these relations themselves.  For this comparison we also calculated  alternative stellar radii directly from the $R(m)$ relation and tabulated them in the row just below the other set of stellar radii (see Table \ref{tbl:simlightcurve}).

In net, we find reasonably good agreement between the results from the simultaneous lightcurve solutions and those found from the physically-based solutions in Sect.~\ref{sec:MCMC}.  The main gain of the new approach described in this section is the much better determination of $\omega_{\rm A,B}$ and $e_{\rm A,B}$ using the simultaneous solutions. Our solution revealed very rapid rates of apsidal advance. For both binaries the yearly precession of the orbital ellipses was found to be about 60\degr. At this point, however, it should be kept in mind that, from this result, it does not follow that one complete revolution of the apsides would take only about six years.  In the next two sections we discuss the dynamical properties and consequences of the common gravitational perturbations in such a tight quadruple system. We find that such short-term effects as, e.g., the periastron passage of the two binaries in their outer orbit around each other, may significantly alter the longer timescale (sometimes called `secular') apsidal advance rates. 
This can lead to large enhancements in apsidal motion on timescales of months or even weeks.

\section{Numerical Simulation of the Orbits}
\label{sec:numerical}

Perhaps the most interesting features of this quadruple system are the large ETVs measured over an interval of only 80 days (see Fig.~\ref{fig:ETVs}).  This clearly points to the two binaries being in a relatively close and interactive orbit.  Since the Keck AO image of `R-S' is unresolved at the 0.05$''$ level, and the distance is estimated to be $\sim$600 pc (Table \ref{tbl:BN}), we already know that the projected size of the outer orbit of the quadruple cannot be more than $\sim$30 AU.  However, the question then arises as to just how close the orbits of the two binaries must be in order to induce the observed level of ETVs (Table \ref{tbl:ETVs}).

We have attacked this question using two different approaches.  In the first, we directly simulate, via numerical integration, a wide range of quadruple systems, each of which contain binaries closely representing A and B whose properties we have determined fairly well (see Sects.~\ref{sec:MCMC} and \ref{sec:lcfactory}).  In the second approach we gain some further insight into the numerical results by the application of a number of analytic approximations to the orbital perturbations.

\begin{figure}[t]
\begin{center}
\includegraphics[width=0.48 \textwidth]{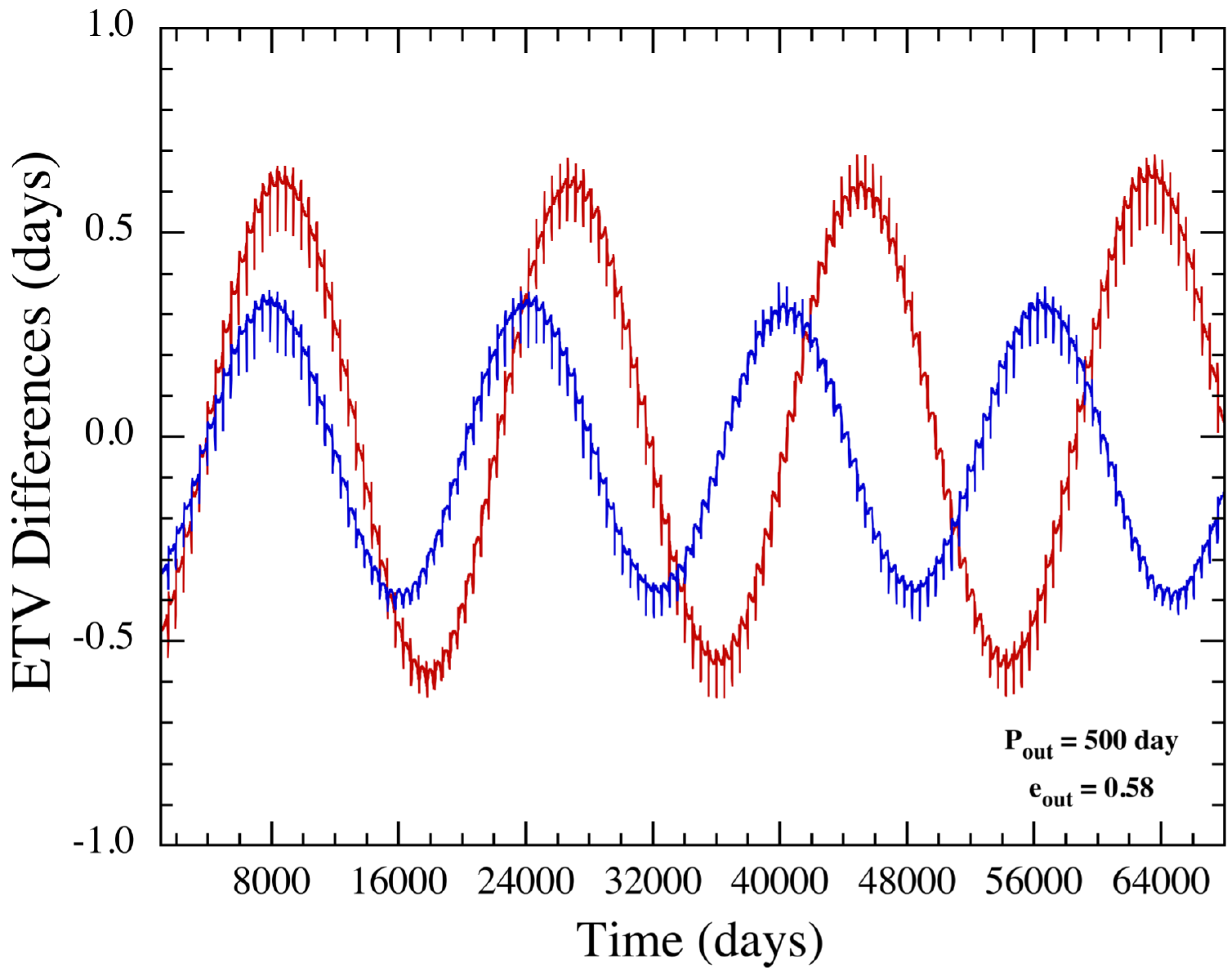}
\includegraphics[width=0.49 \textwidth]{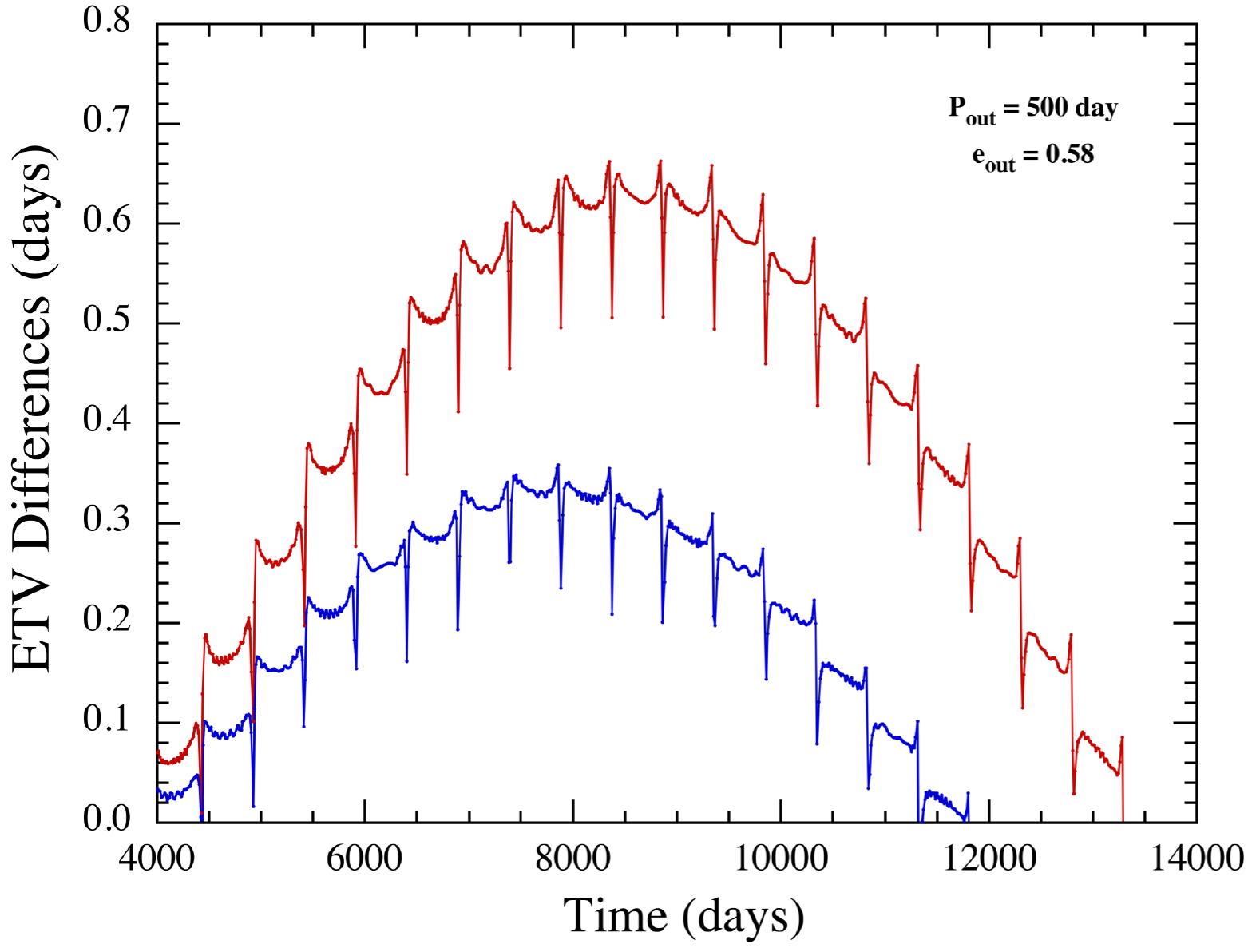}
\caption{Simulated ETV curves for an illustrative outer orbit of binary A around binary B with $P_{\rm orb} = 500$ d and $e_{\rm out} = 0.58$. The red and blue curves are the ETV {\em differences} between the primary and secondary eclipses for the A and B binaries, respectively (with the mean difference $P_{\rm bin}/2$ subtracted).  We plot the ETV differences because that is what the relatively short K2 observations are able to measure.}
\label{fig:sim_ETVs}
\end{center}
\end{figure}  

\begin{figure}[h]
\begin{center}
\includegraphics[width=0.476 \textwidth]{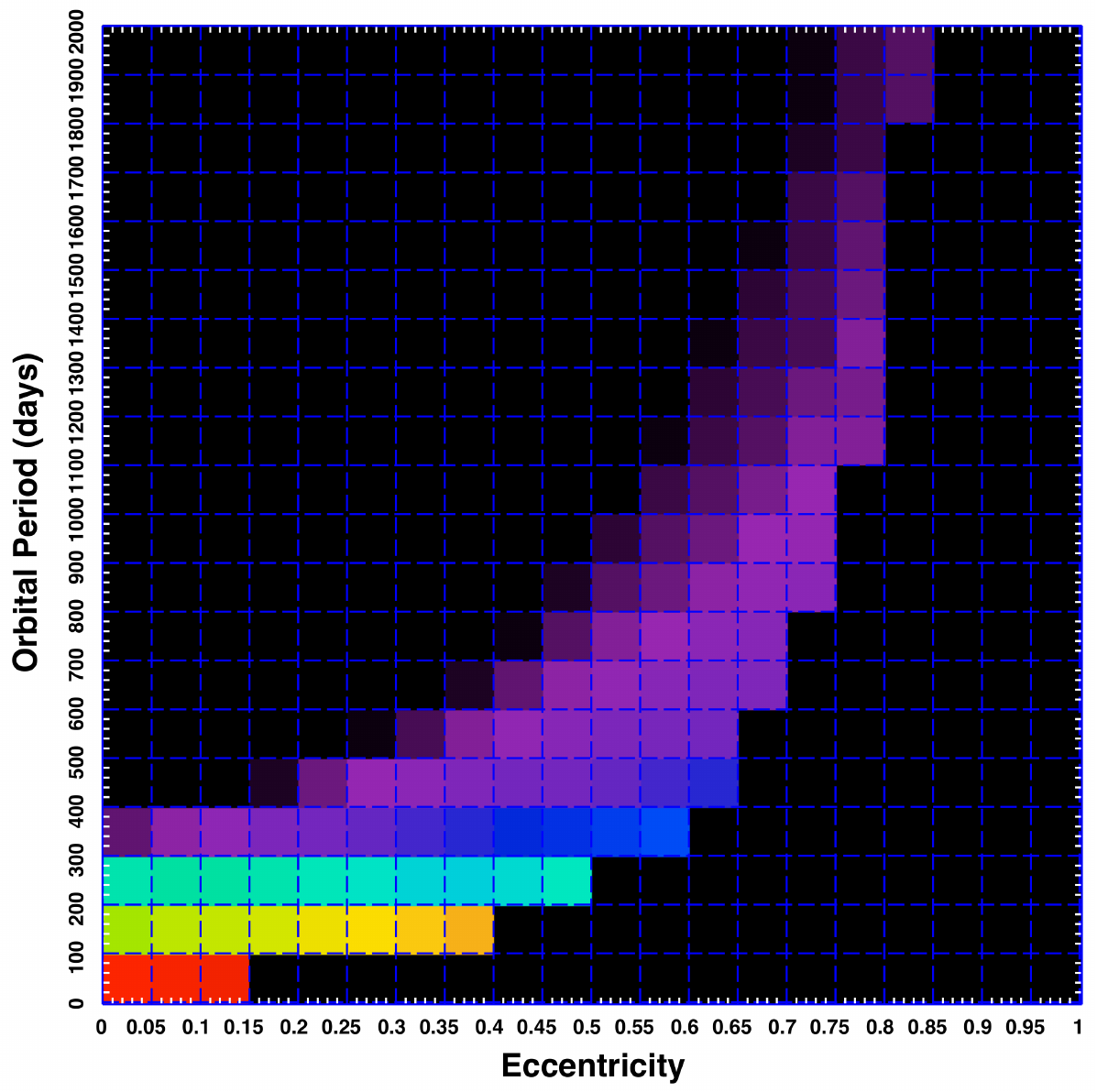} 
\caption{Exploration of the $P_{\rm out}-e_{\rm out}$ plane to determine the importance of dynamically driven apsidal motion in the EPIC 220204960 quadruple system.  The colors indicate the fraction of time during $P_{\rm out}$ when the induced ETVs over the course of 80 days match or exceed those observed during the K2 observations.  Red, orange, cyan, blue, purple, and dark purple represent 90\%, 80\%, 40\%, 25\%, 10\%, and 1\%, respectively. }
\label{fig:ETV_image}
\end{center}
\end{figure}  

For the numerical integrations of the quadruple orbit, we started with binary A and binary B (of known properties; see Table \ref{tbl:MCMC} and \ref{tbl:simlightcurve}) in an outer orbit whose parameters we choose from a grid. The basic 2D grid parameters are: (1) the outer orbital period, $P_{\rm out}$ and (2) its eccentricity, $e_{\rm out}$.  The known masses of the two constituent binaries then determine the semimajor axis of the quadruple system.  Motivated by the near 90$^\circ$ orbital inclination angles of the two individual binaries, we arbitrarily took the mutual inclination between the two binaries to be 0$^\circ$ (a reasonable, but still unproven, assumption).  We also assume that the inclination of the outer orbit with respect to us on the Earth is 90$^\circ$, but since our observation is not long enough to observe eclipses of binary A by binary B, or vice versa, this latter assumption is mostly immaterial.  The initial value of the outer argument of periastron, $\omega_{\rm out}$ was simply taken to have an arbitrary value because (i) we follow the system for many outer orbits during which time the quadruple system can precess, and (ii) the interactions in the binary are not materially dependent on $\omega_{\rm out}$ so long as we record our numerical results over a number of complete cycles of $P_{\rm out}$.  
	
The grid of outer orbits we covered ranged from $P_{\rm out} =100$ days to 2000 days, in steps of 100 days, and $e_{\rm out} = 0 - 1$ in steps of 0.05.  All orbits were integrated for a total duration of 200 years.  We used a simple Runge-Kutta 4th order integrator with a fixed timestep of 4 minutes.  The eclipse times were interpolated to an accuracy of a few seconds.  

We did limit the grid of outer orbits to values of $P_{\rm orb}$ and $e_{\rm out}$ that would be long-term dynamically stable according to the criterion\footnote{Here we are using the Eggleton \& Kiseleva (1995) criteria for 3-body dynamical stability for our 4-body problem by treating each binary, in turn, as a point perturber for the other.} of Eggleton \& Kiseleva (1995):
\begin{equation}
P_{\rm out} \gtrsim 5.0 \,P_{\rm bin} \frac{ \left(1+e_{\rm out} \right)^{3/5}}{ \left(1-e_{\rm out}\right)^{9/5}}
\end{equation}	
where we have taken the mass ratio between binary A and binary B to be unity.  Inadvertently, we did attempt to integrate a couple of systems that were just somewhat beyond this stability line, and those systems indeed disintegrated.  

During the course of each orbital simulation we kept a tabulation of the eclipse times of both the primary and secondary eclipses, including all physical and light travel time delays.  Because the 80-day K2 observation is so relatively short, we were able to measure only a linear `divergence' of the ETVs of the primary eclipse relative to the secondary eclipse.   Accordingly, in the numerical simulations of the orbits, we also tabulated the differences in ETVs between the primary and the secondary.  An illustrative example of these ETV differences is shown in Fig.~\ref{fig:sim_ETVs}.  The top panel shows the ETV differences over the course of approximately 180 years for the A binary in red, and the B binary in blue.  The assumed values of $P_{\rm out}$ and $e_{\rm out}$ for this example were 500 days and 0.58, respectively.  The large sinusoidal features are the approximately 50-year apsidal motion of the binaries.  A zoomed-in view of the ETV differences are shown in the bottom panel of this same figure.  The large dips in the curve every 500 days are due to the periastron passage of the two binaries in their outer orbit when the mutual interactions are the highest.

What we would like to extract from diagrams like this are the changes in ETV differences from eclipse to eclipse.  Even more important is the maximum ETV difference that can accumulate over an 80-day interval that matches the K2 observations.  Thus, for each quadruple whose ETVs are followed for 200 years, we record how often the ETV differences over the course of 80 days exceed those that are observed (Table \ref{tbl:ETVs}), and for what fraction of the outer orbital cycle.  

We summarize these results of our numerical integrations of quadruple orbits in Fig.~\ref{fig:ETV_image}.  The grid shown in the figure covers $P_{\rm out}$ from 100 - 2000 days in steps of 100 days, and $e_{\rm out}$ from 0 to 1 in steps of 0.05.  The color coding of the image display represents the fraction of time that the ETVs match or exceed those observed from EPIC 220204960 over an 80-day interval with K2.  Red, orange, cyan, blue, and purple correspond to fractions of the time exceeding 90\%, 80\%, 40\%, 25\%, and 10\%, respectively. The faintest purple regions are indicative of the fact that such large ETVs would be rare, i.e., occur $\lesssim 1\%$ of the time.  Systems to the left of this colored region will exhibit such large ETVs either extremely rarely, or not at all.  Systems to the right of the colored region are unstable.

\begin{figure}
\begin{center}
\includegraphics[width=0.48 \textwidth]{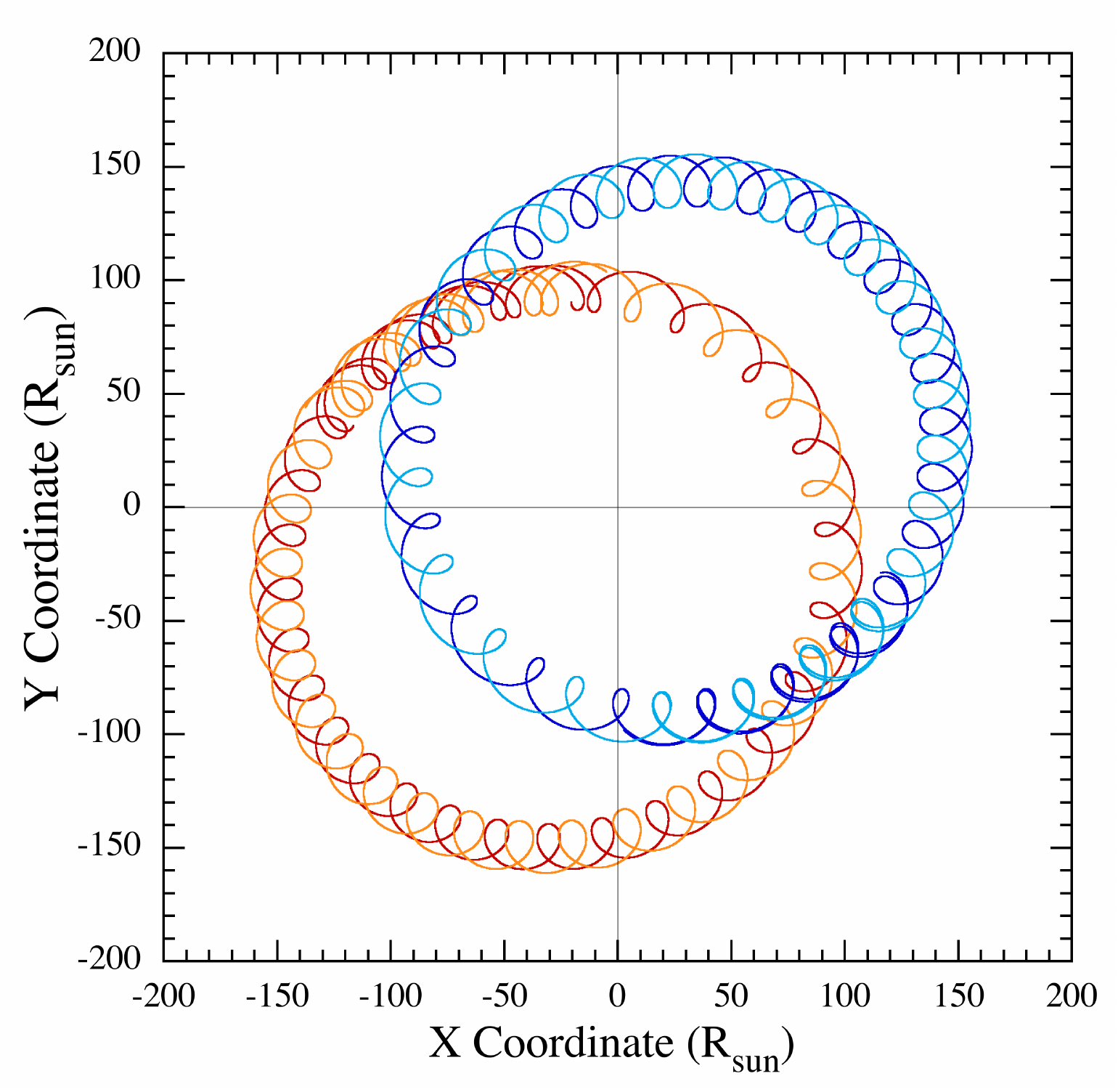}
\caption{Orbital motion of the EPIC 220204960 quadruple system for an assumed illustrative outer orbital period of 500 days and eccentricity of 0.3. The orbital tracks of all four stars are shown in different colors. }
\label{fig:orbits}
\end{center}
\end{figure}  

We conclude from this study that it is most probable that the outer orbit in this system has $P_{\rm out}$ in the range of $300-500$ days.  It is plausible that $P_{\rm orb}$ could be as long as $2-4$ years, but then we would have to have been extremely lucky to see the large ETVs exhibited by both binaries.  Finally, we show in Fig.~\ref{fig:orbits} a tracing of the four stars in their binary and quadruple orbits for the illustrative case of $P_{\rm out} = 500$ days, and $e_{\rm out} = 0.3$.  

\section{Analytic Assessment of the Outer Orbit}
\label{sec:analytic}

In order to gain some analytic insight into the ETVs that one binary induces in the other, we treat each binary as a point perturber for the other. We have good reasons for supposing that both the inner (i.e., binary) orbits and also the outer (quadruple) orbit are coplanar.  If this were not so, and at least one of the two binary orbits were tilted with respect to the outer orbit, dynamical interactions would drive orbital precession for both the tilted binary, as well as the outer orbit.  Therefore, even the other binary orbit would no longer be coplanar with the outer orbit.  As a consequence all three orbits would precess continuously.  In such a scenario we would have to be extremely lucky to observe eclipses in both binaries at the same time. Thus, a more probable possibility is that all three orbits should be (nearly) coplanar.  For such a configuration, we need only consider the analytic forms of the perturbations for the strictly coplanar case.

As is known (see, e.g., Brown 1936), hierarchical triples exhibit periodic dynamical perturbations on three different timescales: $\sim P_\mathrm{in}$, $\sim P_\mathrm{out}$ and $\sim P_\mathrm{out}^2/P_\mathrm{in}$. We omit the smallest amplitude shortest timescale ones, and consider only the other two groups.

First we turn to the longest (sometimes called as ``apse-node'') timescale perturbations. In the framework of the quadrupole-order, hierarchical, three-body approximation for coplanar orbits, the apsidal precession rate is a pure, algebraic sum of the relativistic, classical tidal, and dynamical (third body) contribution, and it is also constant in time apart from low-amplitude fluctuations on the timescales of the other two, shorter-period-class perturbations. Therefore, in this scenario, the rate of apsidal advance of the inner binary can be written as:
\begin{equation}
\dot \omega_\mathrm{in}=\dot \omega_\mathrm{GR}+\dot \omega_\mathrm{tidal}+\dot \omega_\mathrm{dyn},
\end{equation}
where the contributions from the first two terms are given by Levi-Civita (1937) and Kopal (1959) for $\dot \omega_{\rm GR}$, and Cowling (1938) and Sterne (1939) for $\dot \omega_{\rm tidal}$.  The dynamical term due to driven precession by the presence of the third body is: 
\begin{equation}
\dot \omega_\mathrm{dyn}=\frac{3\pi}{2} \frac{M_\mathrm{out}}{M_\mathrm{in}+M_\mathrm{out}} \frac{P_\mathrm{in}}{P_{\rm out}^2} \frac{(1-e_\mathrm{in}^2)^{1/2}}{(1-e_{\rm out}^2)^{3/2}}
\label{eqn:domdyn}
\end{equation}
(e.g., Mazeh \& Shaham 1979).
where $M_\mathrm{in}=m_1+m_2$ is the total mass of the inner binary, while $M_\mathrm{out}$ is the mass of the third component, which in our case is the total mass of the perturbing, other binary system. 

We have evaluated $\dot \omega_\mathrm{GR}$, $\dot \omega_\mathrm{tidal}$, and $\dot \omega_\mathrm{dyn}$ for a reasonable set of system parameters, and the results are given in Table \ref{tbl:apsidalmotion}.  As one can see, the apsidal motion in both binaries is highly dominated by the dynamical perturbations of the other binary and therefore, both the relativistic and tidal contributions can safely be neglected.

\begin{table}
\centering
\caption{Apsidal Motion Properties of the Quadruple Stars}
\begin{tabular}{llcc}
\hline
\hline
\multicolumn{2}{c}{Parameter} & Binary A & Binary B  \\
\hline
$\dot \omega_\mathrm{GR}$ & [rad/day] &$1.03\times10^{-7}$& $0.93\times10^{-7}$\\
& [$''$/yr] & 7.77 & 7.00\\
$\dot \omega_\mathrm{tidal}$ & [rad/day] &$2.40\times10^{-9}$& $1.87\times10^{-9}$\\
& [$''$/yr] & 0.18 & 0.14\\
$\dot \omega_\mathrm{dyn}^a$ & [rad/day] &$2.37\times10^{-4}$& $2.44\times10^{-4}$\\
& [deg/yr] & 4.97 & 5.11\\
$\dot \omega_\mathrm{dyn}^b$ & [rad/day] &$3.21\times10^{-5}$& $3.30\times10^{-5}$\\
& [deg/yr] & 0.67 & 0.69\\
$P_\mathrm{apse}^a$& [yr] & 72.5 & 70.4\\
$P_\mathrm{apse}^b$& [yr] & 535 & 519 \\
\hline 
\end{tabular}
\label{tbl:apsidalmotion}

{\bf Notes.} $a$: For assumed parameters: $P_\mathrm{out}=500$\,d and $e_\mathrm{out}=0.58$; $b$: $P_\mathrm{out}=1000$\,d, $e_\mathrm{out}$=0.0.
\end{table}

In what follows, instead of discussing the ETVs occurring in the primary and secondary eclipses separately, we concentrate on the {\em difference} between ETVs of the primary and secondary eclipses. Such a treatment is quite appropriate in those cases where the observing window is much shorter than the period of the ETVs. The subtraction of the primary ETV from that of the secondary eclipse results in terms which have similar signs for the primary and secondary ETVs formally vanishing.  The only remaining terms are those which anticorrelate between the primary and secondary ETV curves. As a result, the usual light-travel time effect, i.e., the R\o mer-delay is automatically eliminated, together with any other incidental period-change mechanisms, that would result in correlated variations in the primary and secondary eclipse timings.

The time displacement between the secondary eclipses and the mid-time between the primary eclipses is (see e.g., Sterne 1939):
\begin{equation}
D=\frac{P}{\pi}\left\{\arctan\left[\frac{e\cos\omega}{\left(1-e^2\right)^{1/2}}\right]+\left(1-e^2\right)^{1/2}\frac{e\cos\omega}{1-e^2\sin^2\omega}\right\},
\label{eqn:Ddef}  
\end{equation}
where we omit the very week inclination dependence (see, e.g., Gim\'enez \& Garcia-Pelayo 1983).  Since both binaries are in low-eccentricity orbits, we can safely use the first order term of the usual expansion of (\ref{eqn:Ddef}) as,
\begin{equation}
D\simeq\frac{P}{\pi}2e\cos\omega+\mathcal{O}(e^3),
\label{eqn:Dexcdef}  
\end{equation}
which, naturally gives back Eqn.~(\ref{eqn:ecom}). In the coplanar case of our quadruple perturbation model, there are no perturbations either in the inner eccentricity, or the anomalistic period and, therefore, for our binaries we find the rate of change in the ETV differences due to apsidal time-scale forced precession to be:
\begin{equation}
\dot{D}_{\rm apse} \simeq-\frac{2P_\mathrm{in}}{\pi}\dot \omega_\mathrm{in}e_\mathrm{in}\sin\omega_\mathrm{in}
\label{eqn:DP1}
\end{equation}
We might next substitute $\dot \omega_{\rm dyn}$ from Eqn.~(\ref{eqn:domdyn}) for $\omega_{\rm in}$ in Eqn.~(\ref{eqn:DP1}), but this will not be necessary as we shall see.

At this point, before trying to compare the observed and theoretical
ETV differences, we must also include the shorter-term, $P_2$-timescale effects.
For the $P_2$-timescale third-body perturbations in the quadrupole-approximation,
the same ETV-difference in the coplanar case can be calculated from Eqns.~(5--11) of Borkovits et al.~(2015): 
\begin{equation}
D_{P_\mathrm{out}} \simeq A_{P_\mathrm{out}}\left(-3e_\mathrm{in} \mathcal{M}\sin\omega_\mathrm{in}+\frac{15}{2} e_\mathrm{in}\mathcal{C}\right)+\mathcal{O}(e_\mathrm{in}^3),
\label{eqn:DP2}
\end{equation}
where
\begin{equation}
A_{P_\mathrm{out}}=\frac{1}{2 \pi} \frac{M_\mathrm{out}}{M_\mathrm{in}+M_\mathrm{out}} \frac{P_\mathrm{in}^2}{P_{\rm out}}\frac{(1-e_\mathrm{in}^2)^{1/2}}{(1-e_{\rm out}^2)^{3/2}}
\label{eqn:defAP}
\end{equation}
and furthermore,
\begin{eqnarray}
\mathcal{M}&=&\phi_{\rm out}(t)-\theta_{\rm out}(t)+e_\mathrm{out}\sin\phi_{\rm out}(t), \\
\mathcal{C}&=&\cos[2\phi_{\rm out}(t)+2\omega_\mathrm{out}-\omega_\mathrm{in}] +e_\mathrm{out} \times  \\
 & & \left\{ \cos[ \phi_{\rm out}(t)+2\omega_\mathrm{out}-\omega_\mathrm{in}] +\frac{1}{3}\cos[3\phi_{\rm out}(t)+2\omega_\mathrm{out}-\omega_\mathrm{in}] \right\} \nonumber
\label{eqn:M_C}
\end{eqnarray}
where $\phi_{\rm out}(t)$ and $\theta_{\rm out}(t)$ are the true and mean anomalies of the outer orbit.   We calculate the temporal variations of these quantities, in accordance with Eqns.\,(54) and (55) of Borkovits et al.~(2011), to obtain:
\begin{eqnarray}
\dot{\mathcal{M}} & \simeq & \frac{2 \pi}{P_{\rm out}} \left[\frac{(1+e_\mathrm{out}\cos\phi_{\rm out})^3}{\left(1-e_\mathrm{out}^2\right)^{3/2}}-1\right]~, \label{eqn:deltaM}\\
\dot{\mathcal{C}} & \simeq & -\frac{4 \pi}{P_{\rm out}} \frac{(1+e_\mathrm{out}\cos\phi_{\rm out})^3}{\left(1-e_\mathrm{out}^2\right)^{3/2}}\sin(2\phi_{\rm out}+2\omega_\mathrm{out}-\omega_\mathrm{in}),
\label{eqn:deltaC}
\end{eqnarray}
where, in the last equation, the much smaller additional terms due to the apsidal advances of both the inner and outer orbits are neglected. 

If we combine all the terms that contribute to the derivative of $D_{P_{\rm out}}$, i.e., $\dot{D}_{P_{\rm out}}$, from Eqns.~(\ref{eqn:DP2}--\ref{eqn:defAP}) and Eqns.~(\ref{eqn:deltaM}--\ref{eqn:deltaC}), we find:
\begin{eqnarray}
\dot{D}_{P_\mathrm{out}}&\simeq& e_\mathrm{in}A_{P_\mathrm{out}} \frac{2 \pi}{P_{\rm out}} \{3\sin\omega_{\rm in} - \frac{(1+e_\mathrm{out}\cos\phi_{\rm out})^3}{\left(1-e_\mathrm{out}^2\right)^{3/2}} \label{eqn:DeltaDP3} \\
&& \times \left[3\sin\omega_\mathrm{in}+15\sin(2\phi_{\rm out}+2\omega_\mathrm{out}-\omega_\mathrm{in})\right] \}, \nonumber 
\end{eqnarray}
where additional small contributions (in the order of $A_{P_\mathrm{out}}\dot{\omega}$) have been neglected. 

Finally, we can add the $\dot{D}_{\rm apse}$ term due to the continuous forced precession of the binaries' orbits found in Eqn.~(\ref{eqn:DP1}) with the $P_2$-timescale dynamical effects from Eqn.~(\ref{eqn:DP2}).  First, however, we express $\dot{D}_{\rm apse}$ in terms of $A_{\rm P_{out}}$: 
\begin{equation}
\dot{D}_{\rm apse} \simeq -\frac{6 \pi}{P_{\rm out}} e_{\rm in} A_{\rm P_{out}} \sin \omega_{\rm in}
\label{eqn:Dapse}
\end{equation}
It is then immediately clear that the $\dot{D}_{\rm apse}$ term cancels with the first term in Eqn.~(\ref{eqn:DeltaDP3}).  Therefore, we find a net difference in the ETVs of the primary and secondary eclipses of:
\begin{eqnarray}
\dot{D}_\mathrm{tot} \simeq \, A_{\rm etv}  \mathcal{F}(\phi_{\rm out})
\label{eqn:deltaDtot}
\end{eqnarray}
where $\mathcal{F}(\phi_{\rm out})$ is given by 
\begin{equation}
\mathcal{F}(\phi_{\rm out})=(1+e_{\rm out} \cos \phi_{\rm out})^3 \left[\sin\omega_\mathrm{in}+5\sin(2\phi_{\rm out}+2\omega_\mathrm{out}-\omega_\mathrm{in})\right]
\label{eqn:F_phi}
\end{equation}
and the dimensionless $A_{\rm etv}$ is defined as:
\begin{eqnarray}
A_{\rm etv} & \equiv & -3 e_{\rm in} \frac{M_{\rm out}}{M_{\rm out}+M_{\rm in}} \left(\frac{P_{\rm in}}{P_{\rm out}}\right)^2 \frac{(1-e_{\rm in}^2)^{1/2}}{(1-e_{\rm out}^2)^3} \nonumber \\
& \simeq & 5 \times 10^{-5} \left(\frac{500 ~{\rm d}}{P_{\rm orb}}\right)^2 \left(\frac{e_{\rm in}}{0.05}\right)\left(1-e_{\rm out}^2\right)^{-3}~d~d^{-1}
\label{eqn:Aetv}
\end{eqnarray}

\begin{figure}
\begin{center}
\includegraphics[width=0.476 \textwidth]{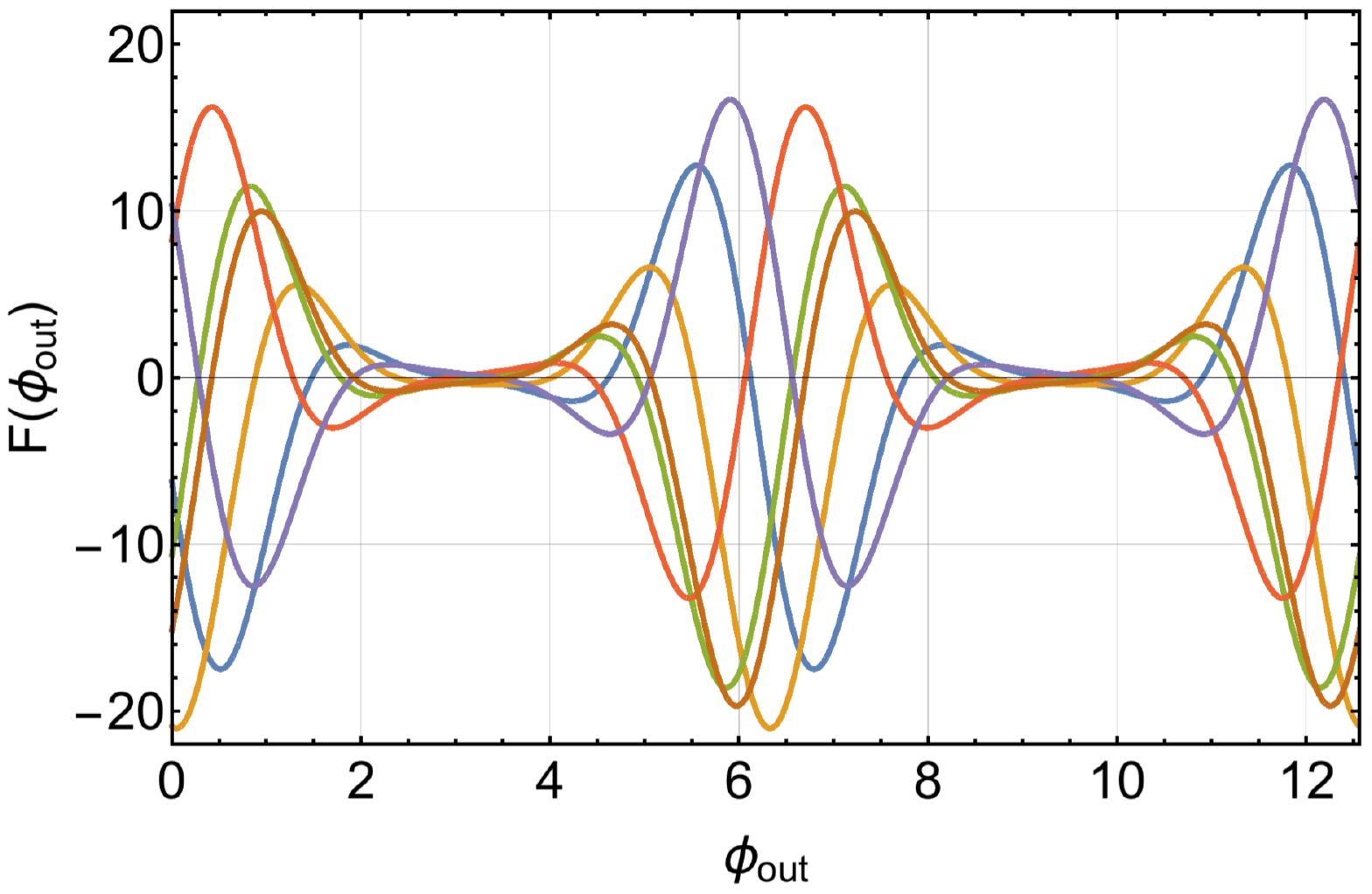} 
\caption{Plots of $\mathcal{F}(\phi_{\rm out})$ for 6 different illustrative values of $\omega_{\rm out}$, where $\phi_{\rm out}$ is expressed in radians.  These curves indicate the dependence of $\dot D_{\rm tot}$ on the true anomaly of the outer orbit according to Eqn.~\ref{eqn:F_phi}. $\dot D_{\rm tot}$ is the difference in the ETVs between the primary and secondary eclipses. }
\label{fig:F_phi}
\end{center}
\end{figure}

The functional part of $\dot{D}_{\rm tot}$, $\mathcal{F}(\phi_{\rm out})$, is plotted in Fig.~\ref{fig:F_phi} for 6 illustrative values of the unknown parameter $\omega_{\rm out}$, and the most likely value of $\omega_{\rm in} \simeq 20^\circ$ or, equivalently, $200^\circ$ (see Table \ref{tbl:simlightcurve}) and with $e=0.58$.  These functions mimic the periodic spikes seen in Fig.~\ref{fig:ETVs} that are evident on the $P_{\rm out}$ timescale.

In conclusion, one can see, that the $P_\mathrm{out}$-timescale perturbations significantly alter the instantaneously measurable ETV difference-variations, and, similarly, the instantaneous apsidal motion rate. We note that this is true even for a circular outer orbit!  In this latter case the first factor in the expression for $\dot{D}_\mathrm{tot}$ in Eqn.~(\ref{eqn:F_phi}) would remain constant, but the second trigonometric term would still result in significant sinusoidal variations.\footnote{Note that even though for $e_\mathrm{out}=0$, $\omega_\mathrm{out}$ loses its meaning, $\phi+\omega_\mathrm{out}$ retains it, and gives the orbital longitude of the third component measured from its ascending node.}

\section{Summary and Conclusions}
\label{sec:concl}

We have presented a quadruple system consisting of a 13.27-day binary orbiting a 14.41-day binary in a quadruple orbit with an outer period that we infer to be about 1 year.  Both binary orbits are slightly eccentric and have inclination angles that are very close to 90$^\circ$.  An adaptive optics image of the host indicates that the current projected separation between the two binaries is $\lesssim 0.05''$, implying a projected physical separation of $\lesssim 30$ AU.  

Because of the relatively wide constituent binaries, the dynamical interactions are quite substantial, larger than for any other known quadruple system (Sect.~\ref{sec:ETVs}).  Indeed, large eclipse timing variations of the order of 0.05 days (over the 80-day observation interval) are detected in both binaries.  As we showed in Sect.~\ref{sec:analytic}, these ETVs are due to a combination of the so-called `physical delay' over the period of the quadruple orbit, and longer-term driven apsidal motion of the mildly eccentric binaries.  

In spite of the faint magnitude of the quadruple system, we were able to obtain radial-velocity-quality spectra at five independent epochs (Sect.~\ref{sec:RVs}).  By carrying out cross-correlation functions against a template M-star spectrum, we are able to see peaks corresponding to all four stars.  After checking all possible combinations of stellar IDs and CCF peaks, we were able to pin down an apparently unique set of star--CCF identifications.  We then carried out orbital fits to these velocities with four free parameters for each binary: $K_1$, $K_2$, $\gamma$, and $\dot \gamma$.  Detection of the acceleration of each binary in its outer orbit seems robust.  The $K$-velocities were then used to determine the stellar masses which are all close to $0.41 \pm 0.05 \, M_\odot$.  

We have analyzed the K2 photometric lightcurve using a physically-based lightcurve emulator to evaluate the binary systems' parameters (Sect.~\ref{sec:MCMC}).  These allow us to make determinations of the four constituent stellar masses that are in good agreement with, and of comparable accuracy to, the RV results. Through this analysis we were also able to measure the orbital inclination angles of the two binaries, as well as to make good estimates of the third-light dilution factors.  
  
Also in regard to the determination of the binary systems' parameters, we re-introduced a technique (to our knowledge used only once before) to analyze the photometric lightcurves of both binaries simultaneously (Sect.~\ref{sec:lcfactory}).  This analysis led to a more robust determination of $\omega$, $\dot \omega$, and thereby a more precise value for the orbital eccentricity for both binaries. 
  
We were able to estimate the period of the outer quadruple orbit via numerical simulations of quadruple systems with constituent binaries of the type we observed in a range of outer orbits covering a grid in $P_{\rm out}$ and $e_{\rm out}$ (Sect.~\ref{sec:numerical}).  After selecting only those quadruple system parameter values that might lead to ETVs of the magnitude we observe, we were led to the conclusion that $P_{\rm out}$ is most likely in the range of 300-500 days.   Analytic estimates of the magnitudes of the expected ETVs are in good accord with the numerical simulations (Sect.~\ref{sec:analytic}).  

Finally, we urge a two-pronged future investigation of this system.  First, it would indeed help define the whole system if interested groups with access to large telescopes could track the radial velocities of these two binaries over an interval of months to a year.  Even 10 RV spectra over the next year might well be sufficient to characterize the outer orbit.  Second, if groups with access to even modest size telescopes could time a few of the eclipses over the next next year, that could also uniquely nail down the outer orbital period. In this regard, we note that if such photometric observations are made in good seeing, where the `B-N' image can be excluded from the aperture, the binary A and B eclipse depths of $\sim$18\% should be relatively easy to measure.

\acknowledgements 

\vspace{0.3cm}

A.\,V.~is supported by the NSF Graduate Research Fellowship, Grant No.~DGE 1144152.  B.\,K. gratefully acknowledges the support provided by the Turkish Scientific and Technical Research Council  (T\"UB\.ITAK-112T766 and T\"UB\.ITAK-B\.IDEP 2219). K.\,P. was supported by the Croatian HRZZ grant 2014-09-8656.  T.\,B. and E.\,F.-D.~acknowledge the financial support of the Hungarian National Research, Development and Innovation Office -- NKFIH Grant OTKA K-113117.  M.\,H.\,K., D.\,L., and T.\,L.\,J.~acknowledge Allan R. Schmitt for making his light curve examining software `LcTools' freely available.  
L.\,N.~thanks NSERC (Canada) for financial support and F. Maisonneuve for his work on the stellar models.
Some of the data presented in this paper were obtained from the Mikulski Archive for Space Telescopes (MAST). STScI is operated by the Association of Universities for Research in Astronomy, Inc., under NASA contract NAS5-26555. Support for MAST for non-HST data is provided by the NASA Office of Space Science via grant NNX09AF08G and by other grants and contracts. 
The MDM Observatory is operated by Dartmouth College, Columbia University, Ohio State University, Ohio University, and the University of Michigan.
A portion of this work was based on observations at the W.~M.~Keck Observatory granted by the California Institute of Technology. We thank the observers who contributed to the measurements reported here and acknowledge the efforts of the Keck Observatory staff. We extend special thanks to those of Hawaiian ancestry on whose sacred mountain of Mauna Kea we are privileged to be guests. 
Some of these results made use of the Discovery Channel Telescope at Lowell Observatory. Lowell is a private, non-profit institution dedicated to astrophysical research and public appreciation of astronomy and operates the DCT in partnership with Boston University, the University of Maryland, the University of Toledo, Northern Arizona University and Yale University. This latter work used the Immersion Grating Infrared Spectrometer (IGRINS) that was developed under a collaboration between the University of Texas at Austin and the Korea Astronomy and Space Science Institute (KASI) with the financial support of the US National Science Foundation under grant AST-1229522, of the University of Texas at Austin, and of the Korean GMT Project of KASI.
Some results are based on data from the Carlsberg Meridian Catalogue 15 Data Access Service at CAB (INTA-CSIC). 

\appendix

\section{Mass-Radius-Temperature Relations for Low-Mass Stars}
\label{app:RTLM}

\subsection{Motivation}
\label{sec:motive}

In Section \ref{sec:MCMC} we used a physically-based lightcurve analysis to infer the constituent masses of the quadruple system.  As part of that analysis we adapted relations for $R(M)$ and $T_{\rm eff}(M)$, where both the radius, $R$, and effective temperature, $T_{\rm eff}$, are assumed to be functions of the mass only (aside of course from the assumed chemical composition).  This is expected to be an excellent approximation for stellar masses $\lesssim 0.6 \, M_\odot$ which will not evolve significantly over a Hubble time. At the opposite mass end, it is good to keep in mind that stars with mass $\lesssim 0.2 \, M_\odot$ will not have fully joined the main sequence for at least 300 Myr (see, e.g., Nelson et al.~1993).

Initially, for the $R(M)$ and $T_{\rm eff}(M)$ relations, we used the analytic fitting formulae for $R(M)$ and $L(M)$ given by Tout et al.~(1996), then solving for $T_{\rm eff}$, and these provided quite reasonable results.  In the case of binary B, both stellar masses are very similar, and therefore we expect a very similar $T_{\rm eff}$ for both stars, and hence similar masses, regardless of the accuracy of the $T_{\rm eff}(M)$ relation.  However, for binary A, since the two eclipse depths are distinctly different (by $\sim$25\%), we can expect that $T_{\rm eff}$ for the two stars will be somewhat different (approximately 6\%).  The difference in mass required to produce this difference in $T_{\rm eff}$ will actually depend sensitively on the {\em slope} of the $T_{\rm eff}(M)$ relation. This is our motivation for re-examining this region of the lower main sequence.

In what follows, we generate a high density of stellar evolution models, and then fit analytic expressions to the results.

\subsection{The Stellar Evolution Code}

All of the stellar models were computed using the Lagrangian-based Henyey method. The original code has been described in several papers (see, e.g., Nelson et al. 1985; Nelson et al. 2004) and has been extensively tested (Goliasch \& Nelson, 2015). The major modifications are due primarily to significant improvements in the input physics that are central to the evolution of low-mass stars and brown dwarfs. In particular, we use the OPAL opacities (Iglesias \& Rogers 1996) in conjunction with the low-temperature opacities of Alexander \& Ferguson (1994), the  Saumon, Chabrier \& Van Horn (1993) equation of state, and the Allard-Hauschildt library of non-gray atmospheres (Hauschildt \& Allard, 1995; Hauschildt et al.~1999). Great care has been taken to ensure that the physical properties blend smoothly across their respective boundaries of validity. Specifically, our treatment enforces continuity of the respective first-order partial derivatives over the enormous range of the independent variables (i.e., density, temperature, and chemical composition) that are needed to fully describe the evolution of low-mass, solar metallicity [Z=0.0173], stars (see Maisonneuve 2007).

\subsection{Results}

We plot in Fig.~\ref{fig:RM} the radius-mass relation from our evolutionary models (at a representative time of 5 Gyr), as bold red circles.  The solid black curve is a fit to a logarithmic polynomial given by the following expression:
\begin{eqnarray}
\log[R(m)] =1.4296 \, \log(m) +1.5792 \, \log^2(m) + 2.8198 \, \log^3(m)+3.0405 \, \log^4(m)+1.2841 \, \log^5(m)
\label{eqn:RM_rel2}
\end{eqnarray}
where $R$ and $m$ are the stellar radius and mass, in solar units, and the logs are to the base 10.  The range of applicability should be limited to $0.1 \lesssim m \lesssim 0.8$.  Overplotted as green circles are the corresponding results of Baraffe et al.~(1998), which are in rather good agreement with our models (see Fig.~\ref{fig:RM}).  The Tout et al.~(1996) $R(m)$ relation (not shown) is also in substantial agreement with the model results, and it has the benefit of working over a much wider range of masses than our expression. The blue circles, with error bars represent 27 well-measured systems as tabulated by Cakirli et al.~(2010),  Kraus et al.~(2011), Carter et al.~(2011b), Dittmann et al.~(2016), and references found therein.  The straight gray line indicates the trend of the data points away from the models.  

We have also fit the {\em empirical} $R(m)$ points in Fig.~\ref{fig:RM} with a function of the same form as in Eqn.~({\ref{eqn:RM_rel2}).  We tested the effect of using this expression on our physically-based lightcurve fits in Sect.~\ref{sec:MCMC}, and we find that it typically yields lower masses for the constituent stars by $\sim$$0.03-0.04 \, M_\odot$ (see the caption to Table \ref{tbl:MCMC}).  However, we do not emphasize these lower masses for two reasons.  First, if anything, the masses determined by the use of Eqn.~(\ref{eqn:RM_rel2}) itself are in better accord with the masses determined from the RV measurements, and second, the vast majority of the empirical masses and radii are from stars in short-period binaries (i.e., with $0.4 \lesssim P_{\rm orb} \lesssim 3$ days) where tidal heating may play a role in enlarging their radii.  

The results for the $T_{\rm eff}$--mass relation for the lower main sequence are shown in Fig.~\ref{fig:TM}.  Again, the red and green circles represent our models in comparison with those of Baraffe et al.~(1998).  The blue circles with error bars are well-measured systems along the lower-mass main sequence.  We fit an analytic expression of the form:
\begin{eqnarray}
T_{\rm eff}(m) =\frac{10^{8.727}\, m^{4.5} + 10^{9.425}\, m^6 -10^{9.928}\, m^7 +10^{9.968}\, m^{7.5}}{1+10^{5.284}\, m^{4.5} + 10^{5.692}\, m^{6.5}}
\label{eqn:TM_rel2}
\end{eqnarray}
to these results, where, again, $m$ is in units of $M_\odot$ and $T_{\rm eff}$ is in K. Note the prediction of a rather flat plateau-like region in $T_{\rm eff}$ over the mass interval $\sim$1/4-1/2 $M_\odot$.  The mass range of applicability for this expression is the same as for Eqn.~(\ref{eqn:RM_rel2}).  Our model points are in good agreement with those of Baraffe et al.~(1998), except near the turnover point at $\sim$0.15 $M_\odot$.  By contrast, the Tout et al.~(1996) $T_{\rm eff}-M$ relation (not shown), while having a somewhat similar shape, is systematically {\em higher} than ours by $\sim$200 K.  This seems likely the result of the Tout et al.~(1996) attempt to fit the entire main sequence (covering 3 orders of magnitude in mass) with a single analytic expression.  There are few good empirical $T_{\rm eff}-M$ pairs over this region, but, if anything, they indicate values of $T_{\rm eff}$ that are $\sim$100-200 K {\em lower} than our analytic relation.

Finally, in Fig.~\ref{fig:RT} we show how our $R - T_{\rm eff}$ relation (deduced by eliminating mass from Eqns.~(\ref{eqn:RM_rel2} and \ref{eqn:TM_rel2}), compares with 21 stars measured interferometrically (taken from data compiled by Newton et al.~2015; interferometric data from Demory et al.~2009; Boyajian et al.~2012).  The region we are concerned with in this work is largely confined to within the purple box.  Aside  from the outlier star (Gl 876) at $R=0.376 \, M_\odot$ and $T_{\rm eff} = 3176$ K, the data are in fairly good agreement with the model curve.

\begin{figure}
\begin{center}
\includegraphics[width=0.62 \textwidth]{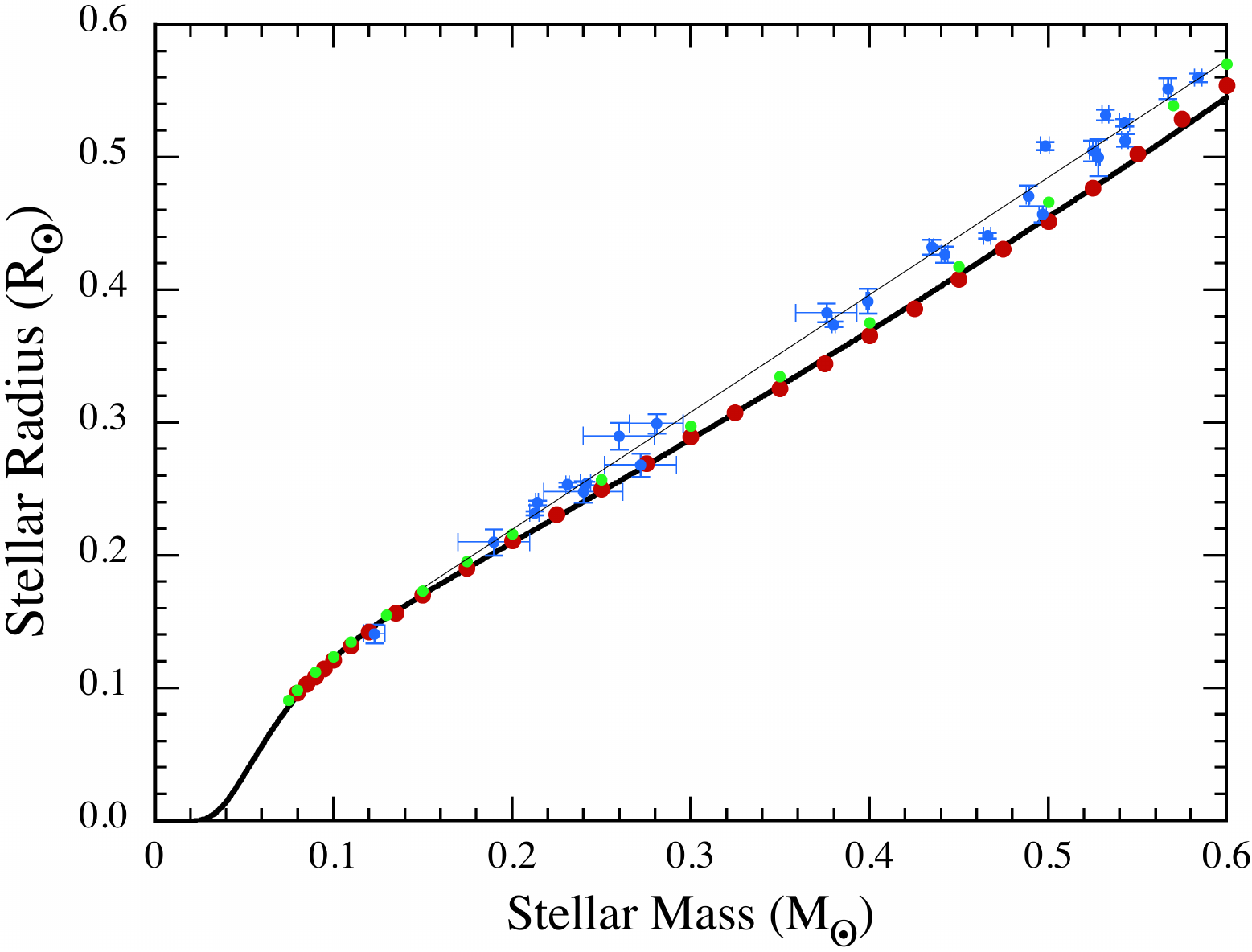}
\caption{Model stellar radius vs.~mass relation on the lower main sequence for solar metallicity stars. The red circles are models that we generated for this work (see text). The light green circles are taken from the Baraffe et al.~(1998) results. The solid black curve is the log-polynomial fit (Eqn.~\ref{eqn:RM_rel2}) to our model points (in red).  Blue circles with error bars are well-measured systems (see, e.g., Cakirli et al.~2010; Kraus et al.~2011; Carter et al.~2011b; Dittmann et al.~2016, and references therein).  The grey straight line marks the trend of the data points away from the models. }
\label{fig:RM}
\end{center}
\end{figure}

\begin{figure}
\begin{center}
\includegraphics[width=0.62 \textwidth]{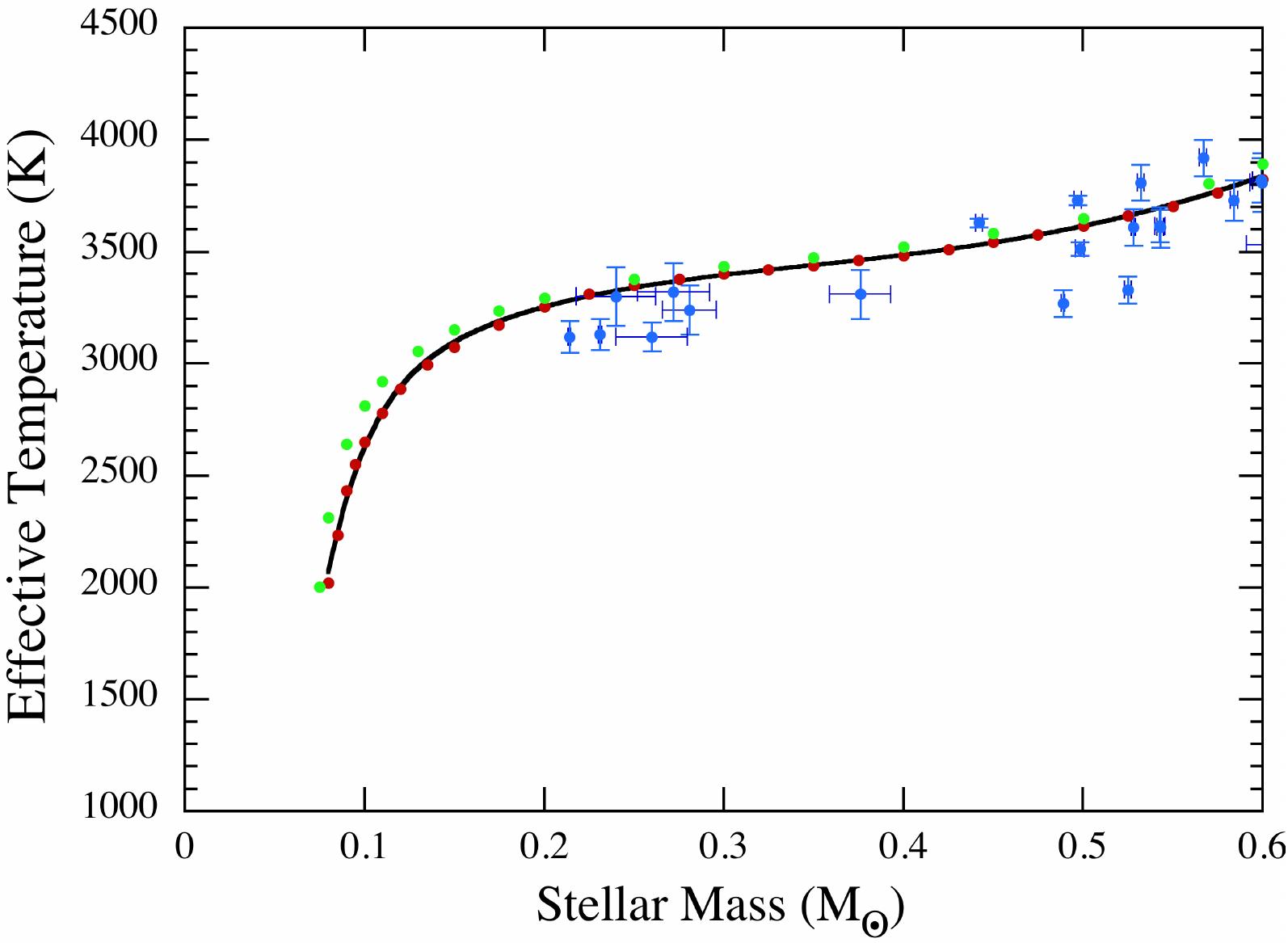}
\caption{Model stellar effective temperature vs.~mass relation on the lower main sequence for solar metallicity stars. The symbols and color coding are the same as in Fig.~\ref{fig:RM}. The solid black curve is the fit of Eqn.~(\ref{eqn:TM_rel2}) to our model points (in red).}
\label{fig:TM}
\end{center}
\end{figure}

\begin{figure}
\begin{center}
\includegraphics[width=0.70 \textwidth]{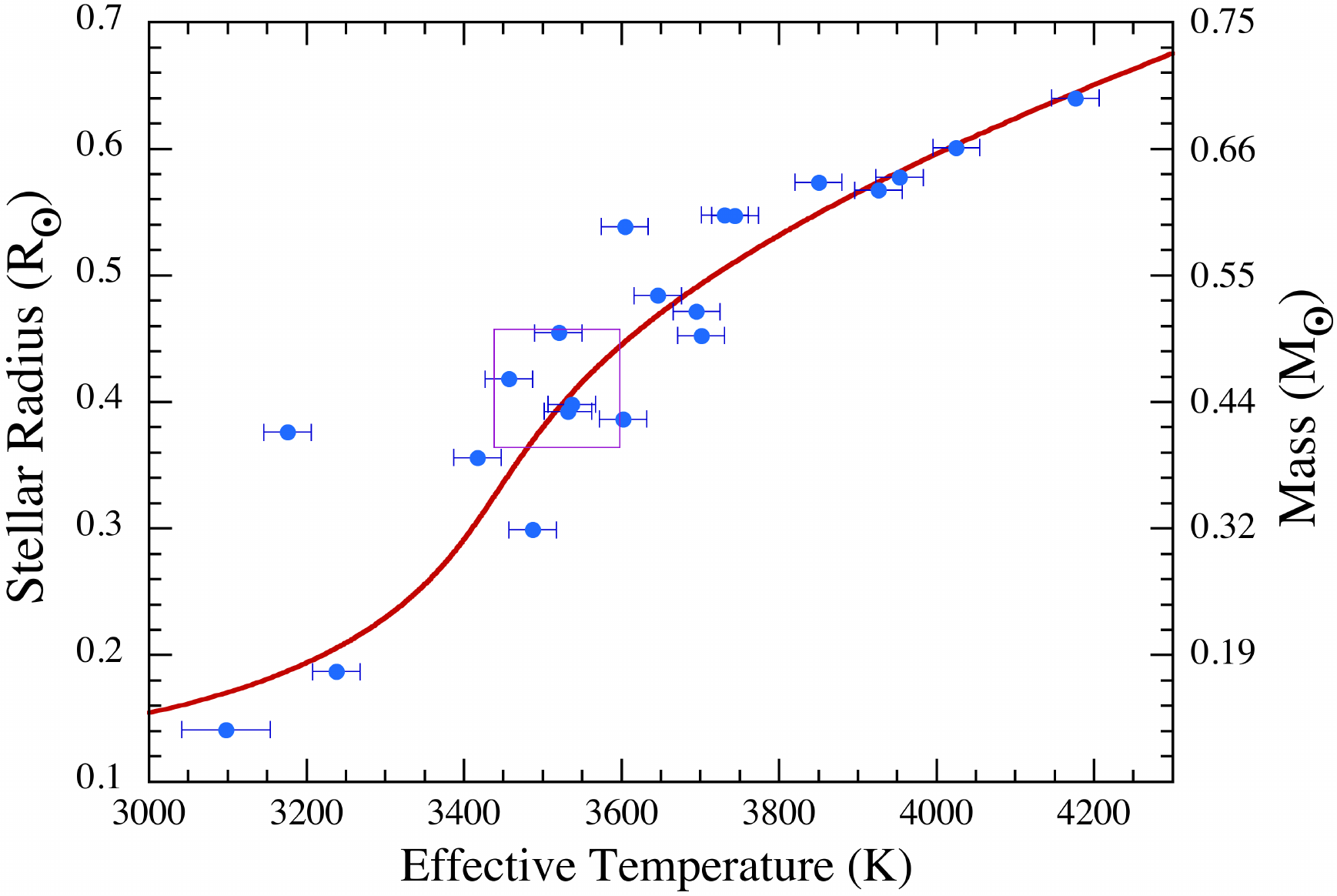}
\caption{Model stellar radius vs.~effective temperature relation on the lower main sequence for solar metallicity stars. The red curve is a parametric expression plotted from Eqns.~(\ref{eqn:RM_rel2}) and (\ref{eqn:TM_rel2}).  The blue circles with error bars are taken from the work of Demory et al.~(2009), Boyajian et al.~(2012), and Newton et al.~(2015). The purple box is the region within which most of our results are derived. The masses listed on the right-hand axis are parametrically inferred from Eqns.~(\ref{eqn:RM_rel2}) and (\ref{eqn:TM_rel2}) and are not measured.}
\label{fig:RT}
\end{center}
\end{figure}

\end{document}